\documentclass[12pt]{article}
\usepackage{a4wide}
\usepackage{graphicx}
\usepackage{amssymb}

\newcommand{\be}{\begin{equation}}
\newcommand{\ee}{\end{equation}}
\newcommand{\ba}{\begin{eqnarray}}
\newcommand{\ea}{\end{eqnarray}}

\newcommand{\order}{\ensuremath{{\cal O}}}
\newcommand{\lapprox}{\lesssim}

\renewcommand{\theequation}{\arabic{section}.\arabic{equation}}

\begin{document}
\begin{titlepage}
\begin{flushright}
UB-ECM-PF 02/01\\
LU TP 02-07\\
March 2002
\end{flushright}
\vfill
\begin{center}
\begin{Large}
Pion and Kaon Electromagnetic Form Factors
 \\[1cm]
\end{Large}
J. Bijnens$^a$ and P. Talavera$^b$\\[2cm]
${}^a$ Department of Theoretical Physics, Lund University,\\ 
S\"olvegatan 14A, S-22362 Lund, Sweden\\
${}^b$ Departament d'Estructura i
    Constituents de la Materia\\
Universitat de Barcelona,
    Diagonal 647,
    Barcelona E-08028 Spain
\\[1cm]
{\bf Pacs:} 12.39.Fe, 12.40.Yx, 12.15.Ff, 14.40.-n\\[0.2mm]

\vfill

{\it Dedicated to the memory of Prof. Bo Andersson}

\vfill

\end{center}

\begin{abstract}
\noindent
We study the electromagnetic
form factor of the pion and kaons
 at low-energies with the use of Chiral Perturbation Theory. The 
analysis is performed within the three flavour framework and
at next-to-next-to-leading order. We explain
carefully all the relevant consistency 
checks on the expressions, present full analytical results for the pion
form factor
and  describe all the assumptions in the analysis.
{}From the phenomenological point of view we make use
of our expression and the available data to obtain the charge
radius of the pion obtaining
$\langle r^2 \rangle_V^\pi=(0.452\pm 0.013)\mbox{fm}^2$,
as well as the low-energy
constant $L_9^r(m_\rho)= (5.93\pm0.43 )\times 10^{-3}$.
We also obtain experimental values for 3 combinations of $\order(p^6)$
constants.
\end{abstract}
\end{titlepage}

\clearpage
\tableofcontents

\section{Introduction}
\renewcommand{\theequation}{\arabic{section}.\arabic{equation}}
\setcounter{equation}{0}

The theory of the strong interaction is a gauge theory of quarks and gluons,
Quantum Chromo Dynamics (QCD).
At high energies and/or short-distances these degrees of freedom and
a perturbation expansion in the gauge coupling constant work well
and provide a lot of experimental tests of it. These provide the basis
for our belief that QCD is indeed the theory of the strong interaction.

At low energies and long distances, it is more difficult to extract
experimental consequences out of QCD. An approach that has
had a lot of successes is the use of chiral symmetry and its constraints
as originally done using current algebra. This relies on the fact
that QCD has a SU(n$_f)_L\times$ SU(n$_f)_R$ chiral symmetry in the limit
of n$_f$ massless flavours. This symmetry is spontaneously broken by
nonperturbative QCD dynamics to the diagonal vector subgroup SU(3)$_V$.

Chiral Perturbation Theory (ChPT) is the effective field theory method to use
this property at low energies. It takes into account the singularities
associated with the Goldstone Boson degrees of freedom caused
by the spontaneous breakdown of chiral symmetry and parametrizes all the
remaining freedom allowed by the chiral Ward identities in
low energy constants (LECs). The LECs are the freedom
in the parts of the amplitudes that depend analytically on the masses
and momenta.
The expansion is then ordered in terms
of momenta, quark masses and external fields. Recent lectures introducing
this area are given in Ref. \cite{chptlectures}. ChPT in its modern
incarnation was founded by Weinberg \cite{weinberg} after earlier work
by Pagels and collaborators \cite{Pagels}. Gasser and Leutwyler extended
and systematized Weinberg's work in a series of papers which caused
the revival of chiral methods \cite{GLannals,GL1}. The use of the external
field method and dimensional regularization allowed a fully chiral invariant
treatment throughout, obviously independent of any parametrization of
the Goldstone Boson fields.

The expansion is ordered in terms of momenta, meson masses
and external fields. We use
here the standard ChPT counting where the quark mass, scalar and pseudoscalar
external fields are counted as 
two powers of momenta. Vector and axial-vector external currents count
as one power of momentum. The Lagrangian can be ordered as
\ba
\label{EffLag}
{\cal L}^{\mbox{\small effective}} &=&  {\cal L}_2 + {\cal L}_4 + {\cal L}_6 +
\cdots
\nonumber\\ 
&=& {\cal L}_2 + \sum_{i=1}^{10} L_i \; O_4^i
+ \sum_{i=1}^2 H_i \; \tilde{O}_4^i
+ \sum_{i=1}^{90} C_i \; O_6^i
+ \sum_{i=91}^{94} C_i \; \tilde{O}_6^i + \cdots\,.
\ea
The index $i$ in ${\cal L}_i$ stands for the chiral power.
The precise form of ${\cal L}_2$ and ${\cal L}_4$
is given in Sect. \ref{definitions} while ${\cal L}_6$ can be found in
\cite{BCE1}.

This expansion can be done both for the case of two or three light flavours,
the number of terms given here correspond to the three light flavour case.
The lowest order, $\order(p^2)$,
in this expansion corresponds to tree level diagrams
with vertices from ${\cal L}_2$, the next-to-leading order,
NLO or $\order(p^4)$, to one-loop diagrams with vertices from ${\cal L}_2$
or tree level diagrams with one vertex from ${\cal L}_4$ and the rest from
${\cal L}_2$. The next-to-next-to-leading order, NNLO or $\order(p^6)$,
has two-loop diagrams, one-loop diagrams with one vertex from ${\cal L}_4$
and tree level diagrams with one vertex from ${\cal L}_6$ or two vertices
from ${\cal L}_4$ and all other vertices from ${\cal L}_2$.
The loop diagrams take all singularities due to the Goldstone Bosons
correctly into account. 
The singularities are the real predictions of ChPT while
the other effects from QCD are in the values of the LECs.
When comparing ChPT and QCD with experiment, it is thus important to
distinguish between the two types of contributions. In QCD one
has to use a scheme where the regularization is consistent with chiral
symmetry. The remnant of the precise definition at short distances
in QCD reflects itself in the precise value of $B_0$, defined below,
and the high-energy constants $H_1, H_2$ and $C_i, i=91,\ldots,94$.

The present situation is such that at lowest and next-to-leading order,
ChPT is quite predictive since we can with relatively few assumptions determine
all parameters and establish relations between various observables.
In the two-flavour case this program has been to a large extent carried
to NNLO as well. In the three light flavour case this is still in progress.
The present manuscript is one more step in the calculation of mesonic
processes to NNLO. 
We present here the full NNLO calculation in three flavour ChPT of
the pion and kaon electromagnetic form factors.
In particular this allows us
to determine one more low-energy parameter $L_9^r$ to this order.
Additional assumptions on the $\order(p^6)$ constants
are at present still necessary
and we will discuss some of these below as well as experimental determinations
of three of them.

The determination of the $L_i^r$ is important since they are the real
consequences of QCD at low energies, distinguishing it from other theories
with the same chiral symmetry pattern. They are also needed to improve the
predictability of ChPT. The previous fits of $L_i^r$ were done at NLO where
the dependence is at most linear and the correlations can thus be undone by
using appropriate linear combinations of observables. In practice we
still have large correlations present in $L_1^r, L_2^r$ and $L_3^r$. At NNLO
the dependence on the $L_i^r$ can become quadratic and it is easier to perform
the fits using MINUIT, taking correlations into account directly.
This what we have done in our earlier work \cite{ABT1,ABT3,ABT2,ABT4} for
$L_i^r, i=1,\ldots,8$. We now add $L_9^r$ to this list. What we found earlier
is that the added correlations at NNLO are present but not very strong,
but that the dependence on the Zweig suppressed constants $L_4^r$ and $L_6^r$
can become
substantial. We have nothing new to add on this question at present but will
return to it in future work. Work at NLO exists, see \cite{Moussallam}
and references therein. Earlier work on the
electromagnetic form factors in ChPT is the NLO work
of \cite{GL2,BC}, the calculation of the nonanalytic parts by
\cite{GM} and the two-flavour calculation at NNLO \cite{BCT}.
Some results for vector form factors at NNLO in three flavours also
exist \cite{Post1,Post2,Post3}.

Here we first give a few definitions of ChPT to fix our notation in Section
\ref{definitions}.
In Sect. \ref{formfactors} we define the vector form factors.
The known NLO results from \cite{GL2,BC} are rederived and presented in Sect.
\ref{nl}. The main new analytical part of this work, the vector
form factors in the three flavour case in the isospin limit, is presented
in Sect. \ref{nnl} and compared as much as possible with the known results
at this order \cite{Post1,Post2,Post3}. 
The data and our ChPT based fits are discussed in Section \ref{datafits}.
The results for $L_9^r$ to NNLO in ChPT accuracy as well as experimental
determinations of 3 combinations of $\order(p^6)$ constants
are in Section \ref{discussion}. This includes resonance estimates
of the $\order(p^6)$ constants needed in Sect. \ref{resonance}
as well as a comparison with predictions for $L_9^r$ in
Sect. \ref{comparison}.
In Sect. \ref{summary} we summarize our conclusions.

We present  some of the lengthier expressions
for the pion form factor
and the integrals we have used in addition to
those we discussed in \cite{ABT1} in the appendices.

\section{Some definitions}
\label{definitions}
\setcounter{equation}{0}

The expressions for the first two terms in the expansion of the
Lagrangian are given explicitly 
by ($F_0$ refers to the 
pseudoscalar decay constant)
\begin{equation}
\label{L2}
{\cal L}_2 = \frac{F_0^2}{4} \{\langle D_\mu U^\dagger D^\mu U \rangle
+\langle \chi^\dagger U+\chi U^\dagger \rangle \}\, ,
\end{equation}
and
\begin{eqnarray}
\label{L4}
{\cal L}_4&&\hspace{-0.5cm} =
L_1 \langle D_\mu U^\dagger D^\mu U \rangle^2
+L_2 \langle D_\mu U^\dagger D_\nu U \rangle
     \langle D^\mu U^\dagger D^\nu U \rangle \nonumber\\&&\hspace{-0.5cm}
+L_3 \langle D^\mu U^\dagger D_\mu U D^\nu U^\dagger D_\nu U\rangle
+L_4 \langle D^\mu U^\dagger D_\mu U \rangle \langle \chi^\dagger U
+\chi U^\dagger \rangle
\nonumber\\&&\hspace{-0.5cm}
+L_5 \langle D^\mu U^\dagger D_\mu U (\chi^\dagger U+U^\dagger \chi ) \rangle
+L_6 \langle \chi^\dagger U+\chi U^\dagger \rangle^2
\nonumber\\&&\hspace{-0.5cm}
+L_7 \langle \chi^\dagger U-\chi U^\dagger \rangle^2
+L_8 \langle \chi^\dagger U \chi^\dagger U 
+ \chi U^\dagger \chi U^\dagger \rangle
\nonumber\\&&\hspace{-0.5cm}
-i L_9 \langle F^R_{\mu\nu} D^\mu U D^\nu U^\dagger +
               F^L_{\mu\nu} D^\mu U^\dagger D^\nu U \rangle
\nonumber\\&&\hspace{-0.5cm}
+L_{10} \langle U^\dagger  F^R_{\mu\nu} U F^{L\mu\nu} \rangle
+H_1 \langle F^R_{\mu\nu} F^{R\mu\nu} + F^L_{\mu\nu} F^{L\mu\nu} \rangle
+H_2 \langle \chi^\dagger \chi \rangle\,,
\end{eqnarray}
while the next-to-next-to-leading order is a rather cumbersome expression
\cite{BCE1}.
The special unitary matrix $U$ contains the Goldstone boson fields
\be
U = \exp\left(\frac{i\sqrt{2}}{F_0}M\right)\,,\quad
M =\left(\begin{array}{ccc}
\frac{1}{\sqrt{2}}\pi^0+\frac{1}{\sqrt{6}}\eta & \pi^+ & K^+\\
\pi^- & \frac{-1}{\sqrt{2}}\pi^0+\frac{1}{\sqrt{6}}\eta & K^0\\
K^- & \overline{K^0} & \frac{-2}{\sqrt{6}}\eta
	 \end{array}\right)\,.
\ee
The formalism we use is the external field method of \cite{GL1}
with $s$, $p$, $l_\mu$ and $r_\mu$ matrix valued scalar, pseudo-scalar,
left-handed and right handed vector external fields respectively.
These show up in
\be
\chi = 2 B_0\left(s+ip\right)\,,
\ee
in the covariant derivative 
\be
\label{covariant}
D_\mu U = \partial_\mu U -i r_\mu U + i U l_\mu\,,
\ee
and in the field strength tensor
\be
F_{\mu\nu}^{L(R)} = \partial_\mu l(r)_\nu -\partial_\nu l(r)_\mu -i 
\left[ l(r)_\mu , l(r)_\nu \right]\,.
\ee
For our purpose it suffices to set
\be
s = 
\left(\begin{array}{ccc}m_u &  & \\ & m_d & \\ & & m_s\end{array}\right)\,,
\quad
l_\mu = r_\mu =  e A_\mu
\left(\begin{array}{ccc}2/3 &  & \\ & -1/3 & \\ & & -1/3\end{array}\right)
\ee
with $e$ the absolute value of the electron charge and $A_\mu$
the classical photon field.

Even if in Eq.~(\ref{L4}) there is no explicit
reference to the high energy regime, the theory depends on it
via the values of the LECs.

\section{The form factors: analytical results}
\renewcommand{\theequation}{\arabic{section}.\arabic{equation}}
\setcounter{equation}{0}

\subsection{Definition}
\label{formfactors}

The most general structure for the on-shell pseudoscalar-pseudoscalar-vector
Green function
is dictated by Lorentz
invariance. With the additional use of
charge conjugation and electromagnetic gauge invariance
one can parametrize the 
pion and kaon electromagnetic matrix elements as
\ba
\label{ff}
\langle \pi^+ (q)\vert j_\mu \vert \pi^+ (p)\rangle& =&
 (q_\mu+p_\mu) F^\pi_V(t)\,,
\nonumber \\
\langle K^+(q) \vert j_\mu \vert K^+ (p)\rangle& =& 
(q_\mu+p_\mu) F^{K^+}_V(t)\,,
\nonumber \\
\langle K^0 (q)\vert j_\mu \vert K^0 (p)\rangle& =& 
(q_\mu+p_\mu) F^{K^0}_V(t)\,,
\ea
with  $t = (q-p)^2$.
The current $j_\mu$ refers to the electromagnetic current of the light flavours
\be
j_\mu = \frac{2}{3} \left(\bar{u}\gamma_\mu u\right) -
\frac{1}{3}\left( \bar{d}\gamma_\mu d +\bar{s}\gamma_\mu s\right) \,.
\ee
The quantities $F_V^\pi, F_V^{K^0}$ and $F_V^{K^+}$
will be referred to hereafter as the vector form factors
or simply the form factors.
They are also defined in the crossed channel 
$ \langle 0 \vert j_\mu \vert M^a(p) M^b(-q)\rangle$. 

Notice that we explicitly neglect here the part of the electromagnetic
current due to the heavier quarks,
\be
j_\mu^{\mbox{\tiny heavy}} =
\frac{2}{3} \left(\bar c\gamma_\mu c+\bar t\gamma_\mu t\right)
-\frac{1}{3}\left(\bar b\gamma_\mu b\right)\,.
\ee
Its effects are suppressed by Zweig's rule and
by the inverse of the heavy mass.

These form factors are near $t=0$ often described by the charge radius,
\be
\langle r^2 \rangle_V^{\pi,K^+,K^0}
= 6 \left.\frac{d}{dt} F_V^{\pi,K^+,K^0}\right|_{t=0}\,,
\ee
and the quadratic term in the expansion,
\be
c_V^{\pi,K^+,K^0}
= \frac{1}{2} \left.\frac{d^2}{dt^2} F_V^{\pi,K^+,K^0}\right|_{t=0}\,.
\ee
Electromagnetic gauge-invariance imposes the constraints
\be
\label{gauge}
F_V^{\pi}(0) = F_V^{K^+}(0) = 1, \quad F_V^{K^0}(0) = 0\,.
\ee
There is no corresponding form factor for $\pi^0$ and the $\eta$.
They vanish due to charge conjugation.

\subsection{Leading and next-to-leading order}
\label{nl}

\begin{figure}[t]
\begin{center}
\includegraphics[width=\textwidth]{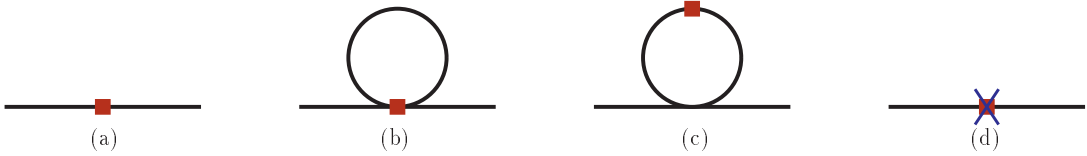}
\end{center}
\caption{One-loop diagrams contributing to the form factors. 
The square stands for the
insertion of the vector external current.
The cross indicates an $\order(p^4)$ vertex from ${\cal L}_4$.
\label{fig1}}
\end{figure}

The leading order expressions for the vector form factors follow from
the tree level diagrams with the Lagrangian (\ref{L2}).
The diagram is depicted in Fig.~\ref{fig1}(a).
The result is rather simple and the form factors are constant, just those
for a point-like particle, and always have the value (\ref{gauge}),
\begin{equation}
\label{tree}
\left.F_{V}^\pi(t)\right|_{p^2} =
\left.F_{V}^{K^+}(t)\right|_{p^2}= 1\,, \quad 
\left.F_{V}^{K^0}(t)\right|_{p^2}= 0\,.
\end{equation}
The simplicity of Eq.~(\ref{tree}) is of great help when computing 
next-to-next-to-leading corrections.
It will be sufficient to use
next-to-leading expressions only when rewriting quark masses and
chiral limit decay constants in terms of the physical meson masses and
decay constants.

Once quantum corrections are taken into account,
loops of pseudoscalars can appear and the Lagrangian ${\cal L}_4$
starts to contribute.
At next-to-leading order we have
the three types of contributions depicted in Fig.~\ref{fig1}, as well as
wave-function renormalization.
The diagrams are:
i) the tree level diagrams giving a polynomial in quark masses
and external momenta whose coefficients depend on the LECs of (\ref{L4})
via the diagram in Fig \ref{fig1}(d).
ii) the tadpole diagrams which contribute a logarithmic dependence on the
quark masses from Fig. \ref{fig1}(b). In the present case these
vanish because of the requirement (\ref{gauge}) when combined with
the contributions from wave-function renormalization.
iii) the unitarity correction due to rescattering shown in Fig. \ref{fig1}(c).

We regulate using dimensional regularization and use
the modified $\overline{\mbox{MS}}$ subtraction scheme as is standard in ChPT
\cite{GLannals,GL1}. 

Our
results are in agreement with those presented in \cite{GL2,BC}
and can be written in the form
\ba
\label{p4}
\left.F^{\pi}_V(t)\right|_{p^4} &=&
\frac{1}{F_\pi^2}\left(
2\,{\cal H}(m_\pi^2,m_\pi^2,t)+{\cal H}(m_K^2,m_K^2,t)\right)\,,
\nonumber\\
\left.F^{K^+}_V(t)\right|_{p^4} &=&
\frac{1}{F_\pi^2}\left(
{\cal H}(m_\pi^2,m_\pi^2,t)+2\,{\cal H}(m_K^2,m_K^2,t)\right)\,,
\nonumber\\
\left.F^{K^0}_V(t)\right|_{p^4} &=&
\frac{1}{F_\pi^2}\left(
-{\cal H}(m_\pi^2,m_\pi^2,t)+{\cal H}(m_K^2,m_K^2,t)\right)\,.
\ea
In terms of the integrals defined in
App. \ref{appintegrals}, the function ${\cal H}$ is defined by
\be
{\cal H}(m^2,m^2,t) =
       \frac{1}{2}\bar{A}(m^2) 
       - \bar{B}_{22}(m^2,m^2,t)
       + \frac{2}{3}\, t L_9^r\,.
\ee
The form factors to $\order(p^4)$ obviously satisfy the relation
\be
\left.F^{\pi}_V(t)\right|_{p^2+p^4} =
\left.F^{K^+}_V(t)\right|_{p^2+p^4}
-\left.F^{K^0}_V(t)\right|_{p^2+p^4}\,.
\ee
At this order in the chiral expansion the expressions for the form factors
depend on a single
LEC, namely $L_9^r$, providing an opportunity for stringent experimental
tests. This was discussed in detail in \cite{GL2} with the then
available data and will be discussed in Sect. \ref{discussion}.

The choice of the pion decay constant $F_\pi$ in (\ref{p4})
is a matter of choice. We could equally well have chosen
$F_0$, the decay constant in the chiral limit, or $F_K$. The difference
is $\order(p^6)$.

\subsection{Next-to-next-to-leading form factors}
\label{nnl}

\begin{figure}[t]
\begin{center}
\includegraphics[width=\textwidth]{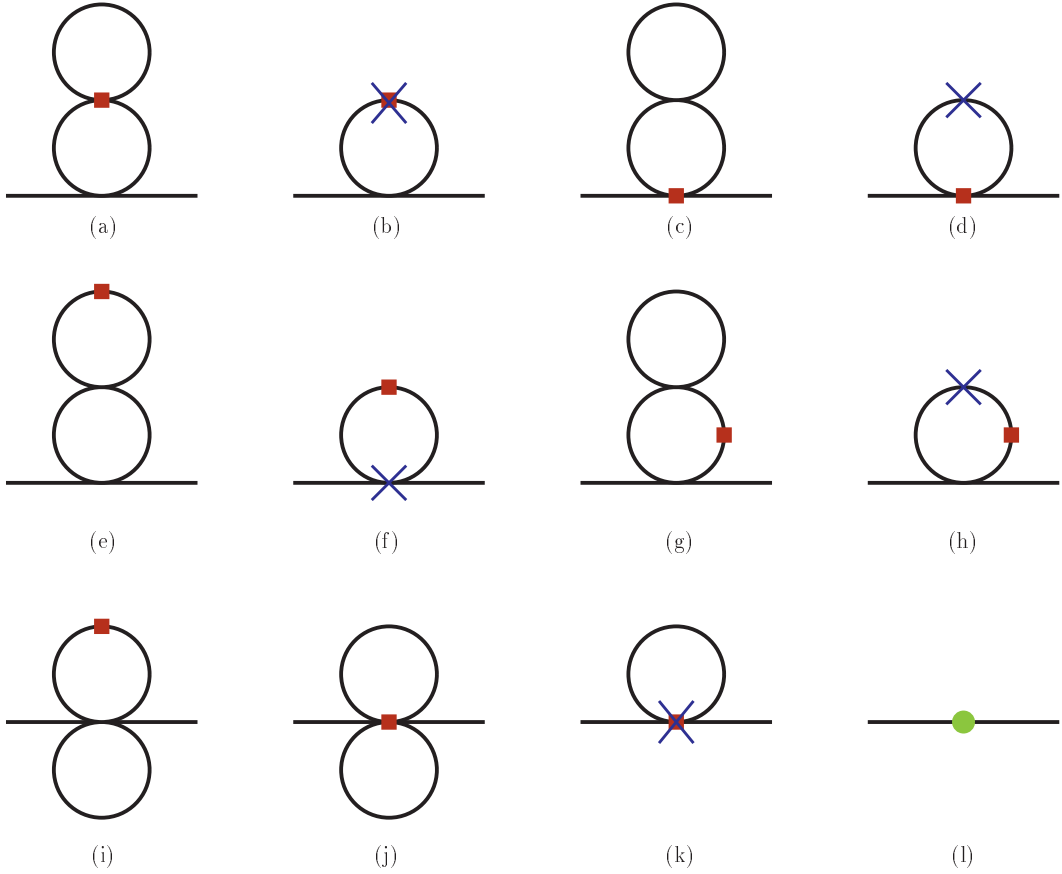}
\end{center}
\caption{The factorizable two-loop diagrams, one-loop diagrams with
vertices from  ${\cal L}_4$, and the tree level diagram
with an ${\cal L}_6$ vertex,
 contributing to the form factors. The
square stands for the
insertion of the vector external current. 
A cross indicates a vertex from ${\cal L}_4$ and a filled circle
a vertex from ${\cal L}_6$.
\label{fig2}}
\end{figure}

\begin{figure}[t]
\begin{center}
\includegraphics[width=\textwidth]{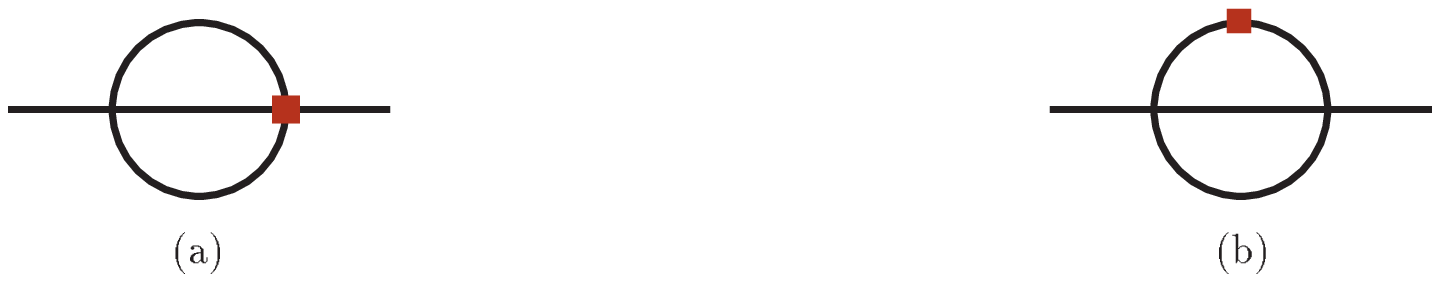}
\end{center}
\caption{Non-factorizable two-loop diagrams contributing to the form factors. 
The square stands for the
insertion of the vector external current. 
There exists also a crossed version of diagram (a), with the
vector current insertion on the left hand side.
\label{fig3}}
\end{figure}

\subsubsection{General techniques and checks}

In order to obtain the expression for the form factors
to $\order(p^6)$ accuracy, one has to
consider the diagrams of Figs.~\ref{fig2} and \ref{fig3} in addition to
wave-function renormalization contributions to the same order. At two-loops
a whole set of new complications surface as compared to one-loop
and especially the regularization and renormalization procedure has to
be dealt more carefully. A long description
tailored to ChPT can be found in Ref. \cite{BCEGS2}, and
issues relevant to definitions of the subtraction are also
treated in \cite{BCE2}.

There are many ways
in which to organize the calculation. One alternative is to
use the one-loop expressions as generalized vertices and propagators
in the two-loop calculation. This shortens the algebra but requires the
knowledge of many one-loop processes also for off-shell external
particles. We chose instead to perform explicitly the calculation
of all Feynman diagrams in terms of lowest order quantities and afterwards
put these into the physical quantities. A number of cancellations
have to occur for both methods to agree and these, described below,
form one of the checks on our calculation.

Our calculations are made in dimensional regularization with the
choice $d = 4 - 2 \epsilon$. In order to include all finite contributions
in $d=4$, it is needed to keep loop integrals and constants in
the diagrams of Fig. \ref{fig2} to order $\epsilon$. In general any function,
$f(m^2_{i0},q^2)$,
appearing in the loop diagrams has a Laurent expansion in $\epsilon$;
\be
\label{expant}
f(m_{i0}^2,q^2) = \frac{1}{\epsilon^2} f^{(-2)}(m_{i0}^2,q^2) 
+ \frac{1}{\epsilon} f^{(-1)}(m_{i0}^2,q^2) 
+ f^{(0)}(m_{i0}^2,q^2) 
+ \epsilon f^{(1)}(m_{i0}^2,q^2)+ \order(\epsilon^2)\,.
\ee
We never encounter terms more divergent than $1/\epsilon^2$
and terms higher order in the expansion do not contribute to the order
we work. The superscripts indicate the power of $\epsilon$ the coefficient
function gets multiplied with.
As an example,
diagram (d) in
Fig.~\ref{fig2} contains two key ingredients: the one-loop two point
function $B$ (see Appendix \ref{appintegrals}) and the 
constants, $L$,  appearing in ${\cal L}_4$, Eq. (\ref{L4}). 
Expanding both of them in Laurent series, one gets the following
naive finite contributions
\begin{equation}
\mbox{diagram (d)}\vert_{\mbox{\tiny finite}} \rightarrow a_1 B^{(-1)} L^{(1)} 
+ a_2 B^{(0)} L^{(0)}  +
a_3 B^{(1)} L^{(-1)}\,,  
\end{equation}
with $a_i$ some polynomial that depends on the masses and/or external momenta.
As shown in \cite{BCE2}, the constants $L^{(1)}$ can be chosen to be
zero, their effect can be absorbed in the values of the $\order(p^6)$
constants $C_i$. We will do so. The contribution containing $B^{(1)}$
can be shown to always have counterparts cancelling it in other diagrams.
This however requires a careful evaluation of all diagrams that makes
this cancellation apparent and in particular, requires careful treatment
of parts of the nonfactorizable diagrams. We have evaluated the latter
in a more general fashion and we are thus obliged to keep
the contributions containing $B^{(1)}$.

{\bf Regularization and renormalization:} 
\label{rere}

In a local field theory, the final divergences are always polynomial
in the masses and momenta. These can thus be subtracted from the freedom
in the choice of parameters in the tree level Lagrangian. We use
here the standard ChPT $\overline{\mbox{MS}}$ subtraction with e.g. for
the coefficients of ${\cal L}_4$
\be
\label{L4sub}
L_i = \left(e^c \mu\right)^{d-4}
\left(\frac{\gamma_i}{16\pi^2(d-4)}+L_i^r(\mu)\right)
\ee
and
\be
c = -\frac{1}{2}\left(\ln(4\pi)+\Gamma^\prime(1)+1\right)\,.
\ee
Notice that Eq. (\ref{L4sub}) implies a choice of $L^{(1)}_i$ as mentioned
above.

To $\order(p^4)$ there are no problems and all divergences occurring
diagram by diagram are polynomial. This is no longer true at $\order(p^6)$
where there are two sources of nonpolynomial divergences. We have the
one-loop diagrams with insertions of ${\cal L}_4$ vertices. The nonpolynomial
part of the loop integrals is multiplied with (\ref{L4sub}) and leads
to divergences proportional to $\ln(M^2/\mu^2)$ and others
with momentum dependence.
The required cancellation of these divergences leads to so called Weinberg
relations. From the one-loop diagrams with ${\cal L}_4$ vertices we
can determine the nonpolynomial divergences 
of the two-loop diagrams. These in turn determine the double poles.

As an example of the above, the $t$ dependent part of the nonlocal divergence
produced by the diagram of Fig. \ref{fig2}(f) must be cancelled by
other diagrams having the same possible divergence,
in this case the diagrams of Fig. \ref{fig2}(a,b,e,g,h,i) and of
Fig. \ref{fig3}(b). The cancellation of both the $t$ dependent and
the $\ln (M^2/\mu^2)$ dependent divergences is a very strong check on our
calculation.

{\bf Double logarithms:} The cancellation as discussed above, can in fact be
used to calculate part of the $\order(p^6)$ correction using only one-loop
diagrams. This has been done for quite a few processes in Ref. \cite{BCE3}.
In the two-flavour case for many quantities these so called double-logarithm
contributions often dominate, in particular they do so for $\pi\pi$-scattering
and the corrections to the masses, \cite{pipishort,BCT} but not
for e.g. the pion form factor \cite{BCT}. 
Our results here agree with the double-logarithm part as calculated
in \cite{BCE3}.

{\bf Terms of the type $L_i^r(\mu)\times L_j^r(\mu)$:}
Similarly to the previous
contribution the $L_i^r(\mu)\times L_j^r(\mu)$
pieces can be obtained in full generality without a complete calculation, since
they only occur via tree diagrams and were also calculated in \cite{BCE3}.
Our results agree with those given there.

{\bf Irreducible two-loop integrals:}
One of the main features of the non-factorizable integrals
at two-loop (see Fig.~(\ref{fig3})) in the three flavour case
is that they can contain different scales even at
zero momentum transfer. 
Those scales are given by the pseudoscalar masses. No exact analytic
expressions are known for all the integrals we need here.
Given the complexity of the integrals,
it is worthwhile to have as many cross-checks on the numerical evaluation
as possible.
We differentiate between the two topologies in Fig.~(\ref{fig3}).
The topology (a) is calculated with
the methods described in \cite{ABT1}. In particular, the subtraction
constants are derived using the methods of \cite{DT,PS}, of which we
obtained an alternative derivation of the recursion relations,
and the dispersive part with a generalization of the methods of \cite{GS}
to different masses. In addition we 
cross-checked the results using the techniques in
\cite{GB,GY}.

For the topology (b) we used the methods given in \cite{GB,GY} and 
cross-checked with \cite{GS} for the case of equal masses.
As an independent check for the tensor structures, we evaluate most of
them numerically and then check that they satisfy the
relations derived using the methods of Passarino and Veltman \cite{PV},
between the various different Lorentz structures.

{\bf Putting the result in physical quantities:}
Up to this level we have computed all the diagrams in terms of the
quantities appearing in the Lagrangian. In particular we have used
the decay constant in the chiral limit and the lowest order meson masses.
Once the amplitude is finite in terms of these bare quantities,
with a cancellation of all nonlocal and nonpolynomial divergences
as well as agreement of the polynomial divergences with the general
evaluation of \cite{BCE2},
we must rewrite these
bare masses and decay constants in terms of the physical ones.
As was mentioned already, the leading order in the 
form factors are mere constants fixed by Ward identities, Eq. (\ref{gauge}).
We thus only need the $\order(p^4)$ expressions for masses and decay constants
to do this rewriting. This is one of the reasons that the expressions
here are so much simpler than in the $K_{\ell4}$ case performed earlier,
allowing us to quote the full result for the pion form factor,
not only a numerical parametrization.

There is of course a certain choice in how this writing in physical
decay constants is done.
At lowest order $F_\pi \equiv F_K \equiv F_\eta = F_0$
but at higher orders nothing prevents us
to write the $\order(p^4)$
expressions with $F_K$ or $F_\eta$, rather than $F_\pi$.

For the masses the situation is easier in this case as well,
the $\eta$ does not appear in the $\order(p^4)$ result, so
the full $\order(p^4)$ result
can be simply written in terms of the pion and kaon mass.
The masses can appear inside the arguments of the loop integrals.
Rewriting those in the physical masses we
expand the function in a Taylor series around the value of the
mass at the next-to-leading order  
\begin{equation}
f(m_{i0}^2;q^2)=f(m_{i}^2;q^2) + \sum_j \left( m_{j0}^2-m_j^2\right)
\frac{\partial}{\partial m_{j0}^2} f(m_{i0}^2;q^2) \vert_{m_{i0} = m_i}\,.
\end{equation}
For instance, for the scalar one and two-point functions, $A$ and $B$,
one obtains
\begin{eqnarray}
A(m_{\pi0}^2) \hspace{-0.2cm}&=&
\hspace{-0.2cm} A(m_\pi^2) + \left(m_{\pi0}^2-m_\pi^2\right)
B(m_\pi^2,m_\pi^2;0)\,, \nonumber\\
B\left(m_{\pi0}^2,m_{K0}^2;q^2\right)
 \hspace{-0.2cm}&=&\hspace{-0.2cm} B(m_\pi^2,m_K^2;q^2)
 +\left( m_{\pi0}^2-m_\pi^2\right) 
C(m_\pi^2,m_\pi^2,m_K^2;q^2) \nonumber\\&&+
\left( m_{K0}^2-m_K^2\right) 
C(m_K^2,m_K^2,m_\pi^2;q^2)\,,
\end{eqnarray}
where the $m_{i0}^2$ are
the bare 
or lowest order quantities and the $m_i^2$
are the equivalent renormalized masses at next-to-leading
order.
Finally the $C$ function is the scalar three point function
\be
C(m_1^2,m_2^2,m_3^2,q^2) =
\frac{1}{i}\int\,d^dp \frac{1}{\left(p^2-m_1^2\right)\left(p^2-m_2^2\right)
\left((p-q)^2-m_3^2\right)}\,.
\ee

Once these steps are performed, we are allowed to identify the masses and
decay constants with the physical ones.
Notice that this identification can be done without any manipulation
in the genuine two-loop topologies from
the very beginning because the difference is of higher order.
As a consistency check, we have
cross-checked that all our final expressions are scale invariant. 
Notice that both the lowest order masses and decay constants and the physical
ones are finite and this rewriting does not introduce any infinities.

{\bf Three-point one-loop integrals:}
There is an extra check on the way that we performed the renormalization
of the masses and decay constants. As is
evident the diagrams (a,b,c,d,g) and (h) in Fig.~(\ref{fig2})
can be obtained by substituting the ``bare propagator'' of
diagrams (b,c) in Fig.~(\ref{fig1}) by a ``dressed'' one-loop one.
In this last method only one and two-propagator loop functions are
permitted. In particular this tells us that the three-point
functions, $C$, appearing naively via diagrams (g) and (h) should
disappear once we promote all bare quantities to the physical
ones.

{\bf Gell-Mann--Okubo relation:}
There is one last comment concerning higher order corrections. At $\order(p^6)$
we have chosen to rewrite all masses, except those inside logarithms and loop
functions, in terms of the pion and kaon mass. The quark masses are written
into those and the $\eta$ mass is removed using the
 GMO relation
in the SU(2) isospin limit, 
\begin{equation}
\label{gmo}
3 m_\eta^2= 4 m_K^2 - m_\pi^2\,.
\end{equation}
As mentioned above already, the vector form factors
do not depend on the eta mass at next-to-leading order,
only at next-to-next-to-leading order. 
This is not necessarily
the case for other quantities, in particular the scalar form factors.
Numerically at low transfer momentum the difference in using or not
Eq.~(\ref{gmo}) for the $\eta$ mass in the remaining loop functions
is small.

{\bf Independence of $\mu$:} A final check on the numerical programs
is that they are independent of the scale $\mu$ when
this scale is varied in the evaluation of the integrals
and the values of the constants are changed appropriately.

{\bf Comparison with previous work:} 
As an independent cross-check of our calculations we compared whenever
 possible with the available expressions in the literature. 
This is the
case for
the $K^0$ form factor which was obtained
up to next-to-next-to-leading order in 
\cite{Post2} for the isospin limiting case. 
We have compared partially our results for the same
process and find agreement with the double logarithm terms and with the pieces 
containing the irreducible integrals.
The same authors considered previously, \cite{Post1},
a certain combination for
the vector form factor that leads
to the Sirlin's relation \cite{Sirlin}. The $q^2$
dependence of this relation is free
of unknown 
constants of $\order(p^6)$
because of the Ademollo-Gatto theorem \cite{ademollo}.
Some expressions for the pion form factor are also in the recent
paper \cite{Post3}.
The analytical comparison with this expression can only be made between the
pieces obtained via the reducible diagrams finding agreement between them.
For the rest of the expression,
irreducible diagrams,
the authors of \cite{Post1,Post3} use like us a numerical approach.
The results of these are in good
agreement.

{\bf FORM:} Most of the algebraic manipulations needed for this work were
performed using FORM 1.0 and
FORM 3.0 \cite{FORM} while independent cross-checks of some structures 
were performed by hand.

\subsubsection{Results}
\label{nnlresults}

We split the full $\order(p^6)$ contribution in several pieces
\be
\left.F_V^P(t)\right|_{p^6}
= \frac{1}{F_\pi^4}\left(
F_{V\mathbf{L}}^P(t)+
F_{V\mathbf{C}}^P(t)+
F_{V\mathbf{B}}^P(t)+
F_{V\mathbf{H}}^P(t)+
F_{V\mathbf{V}}^P(t)
\right)\,,
\ee
with the following definitions:\\
\begin{tabular}{cl}
$F_{V\mathbf{L}}^P(t)$:& The part which depends on the $L_i^r$.\\
$F_{V\mathbf{C}}^P(t)$:& The polynomial dependence on the ${\cal L}_6$
parameters $C_i^r$.\\
$F_{V\mathbf{B}}^P(t)$:& The part with the loop integrals that can be done
analytically.\\
$F_{V\mathbf{H}}^P(t)$:& The part that depends on the sunset integrals.\\
$F_{V\mathbf{V}}^P(t)$:& The part containing the irreducible two-loop
vertex integrals.\\
\end{tabular}\\
The last three contain the proper two-loop diagrams and are the hardest
part of the calculation. The precise separation between them is somewhat
dependent on how the set of nonfactorizable two-loop integrals is
chosen. In addition, there are rather large numerical cancellations between
these three parts because the method to separate factorizable from
nonfactorizable contributions often induces large terms in both which
afterwards cancel.

We have not performed the separation into tadpole and unitarity corrections
here. Because of the requirement (\ref{gauge}) the tadpoles cancel to
a very large extent and the separation between tadpole and unitarity
contributions is in any case representation dependent.

The dependence on the $L_i^r$ is fairly simple:
\ba
\label{appll}
   F_{V\mathbf{L}}^\pi(q^2) &=&
 8 m_\pi^2 \left( 2 L_4^r + L_5^r \right) \bar{A}(m_\pi^2) 
 + 4 m_\pi^2 L_5^r \bar{A}(m_K^2) +  q^2 L_9^r (6
 \bar{A}(m_\pi^2) + 3\bar{A}(m_K^2) ) \nonumber\\&&
 + \bar{B}_{22}(m_\pi^2,m_\pi^2;q^2)   
\left\{  - 16 m_\pi^2 (2 L_4^r + L_5^r ) + 4 q^2 (4 
 L_1^r - 2 L_2^r + 2 L_3^r - L_9^r ) \right\} \nonumber\\&&
 + \bar{B}_{22}(m_K^2,m_K^2;q^2)   \left\{  - 8 m_\pi^2 L_5^r + 4 q^2 L_3^r - 2
   q^2 L_9^r \right\}\,,
\ea

\ba
\label{akpll}
   F_{V\mathbf{L}}^{K^+}(q^2) &=&
 \left(16 m_K^2  L_4^r + 8 m_\pi^2 L_5^r\right)\bar{A}(m_K^2) 
 + 4 m_\pi^2 L_5^r \bar{A}(m_\pi^2) +  q^2 L_9^r (5
 \bar{A}(m_\pi^2) + 4\bar{A}(m_K^2) ) 
\nonumber\\&&
 + \bar{B}_{22}(m_K^2,m_K^2;q^2)   
\left\{  - 32 m_K^2  L_4^r - 16 m_\pi^2 L_5^r  + 4 q^2 (4 
 L_1^r - 2 L_2^r + 2 L_3^r - L_9^r ) \right\}
\nonumber\\&&
 + \bar{B}_{22}(m_\pi^2,m_\pi^2;q^2)   \left\{  - 8 m_\pi^2 L_5^r + 4 q^2 L_3^r - 2
   q^2 L_9^r \right\}
\nonumber\\&&
16 L_5^r L_9^r q^2\left(m_\pi^2-m_K^2\right)\,,
\ea
and
\ba
\label{akoll}
   F_{V\mathbf{L}}^{K^0}(q^2) &=&
  \left(4 m_\pi^2 L_5^r+q^2 L_9^r\right)\,
\left( - \bar{A}(m_\pi^2)+ \bar{A}(m_K^2)\right)
\nonumber\\&&
 + \bar{B}_{22}(m_\pi^2,m_\pi^2;q^2) 
  \left\{   8 m_\pi^2 L_5^r - 2 q^2 (2L_3^r -  L_9^r) \right\}
\nonumber\\&&
 + \bar{B}_{22}(m_K^2,m_K^2;q^2)   
\left\{ -8 m_\pi^2 L_5^r  + 2 q^2 ( 
  2 L_3^r - L_9^r ) \right\}\,.
\ea

\begin{figure}[t]
\begin{center}
\includegraphics[width=0.9\textwidth]{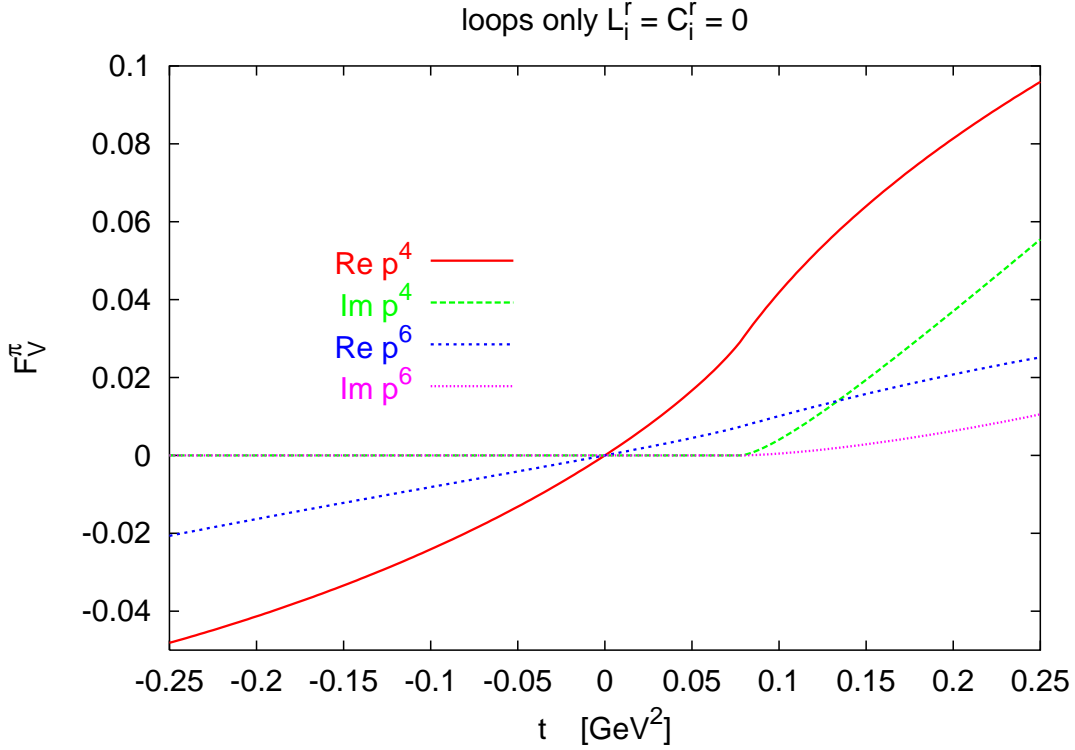}
\end{center}
\caption{The real and imaginary parts of the loop diagrams
 at $\order(p^4)$ and $\order(p^6)$ with all $L_i^r=0$,
for the pion form factor. Notice that there is convergence
both for the real and the imaginary part.
\label{figloops1}}
\end{figure}

\begin{figure}[t]
\begin{center}
\includegraphics[width=0.9\textwidth]{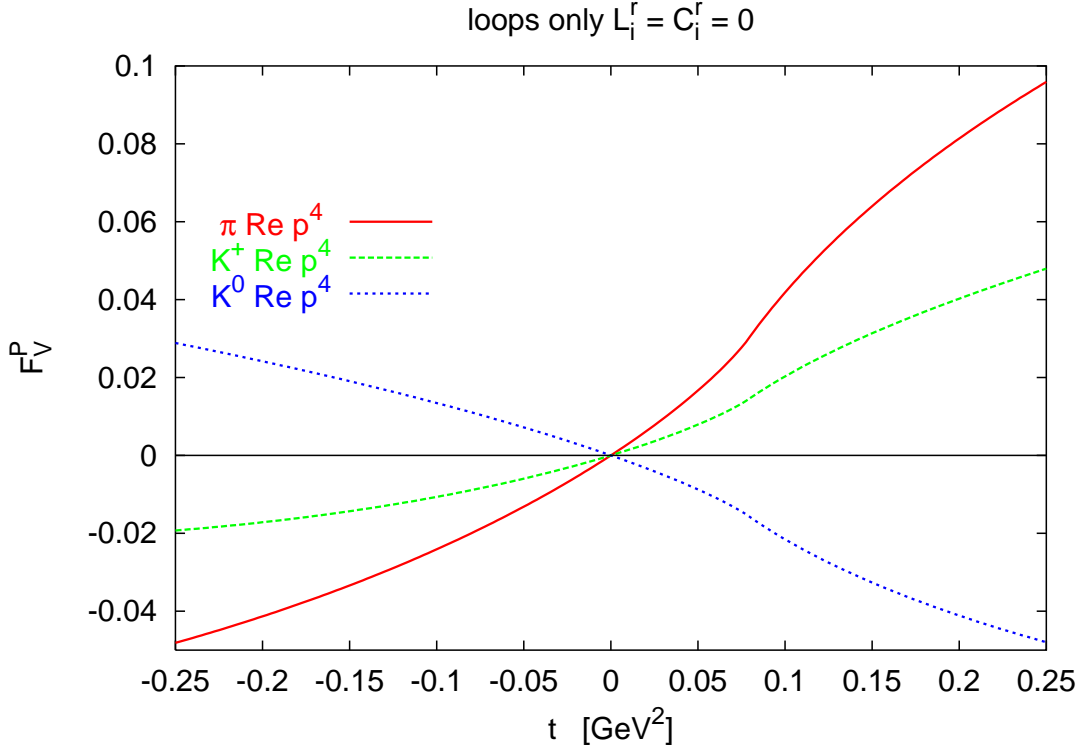}
\end{center}
\caption{The real parts of the loop diagrams
 at $\order(p^4)$ 
for the pion, charged kaon and neutral kaon  form factor.
\label{figloops2}}
\end{figure}

\begin{figure}[t]
\begin{center}
\includegraphics[width=0.9\textwidth]{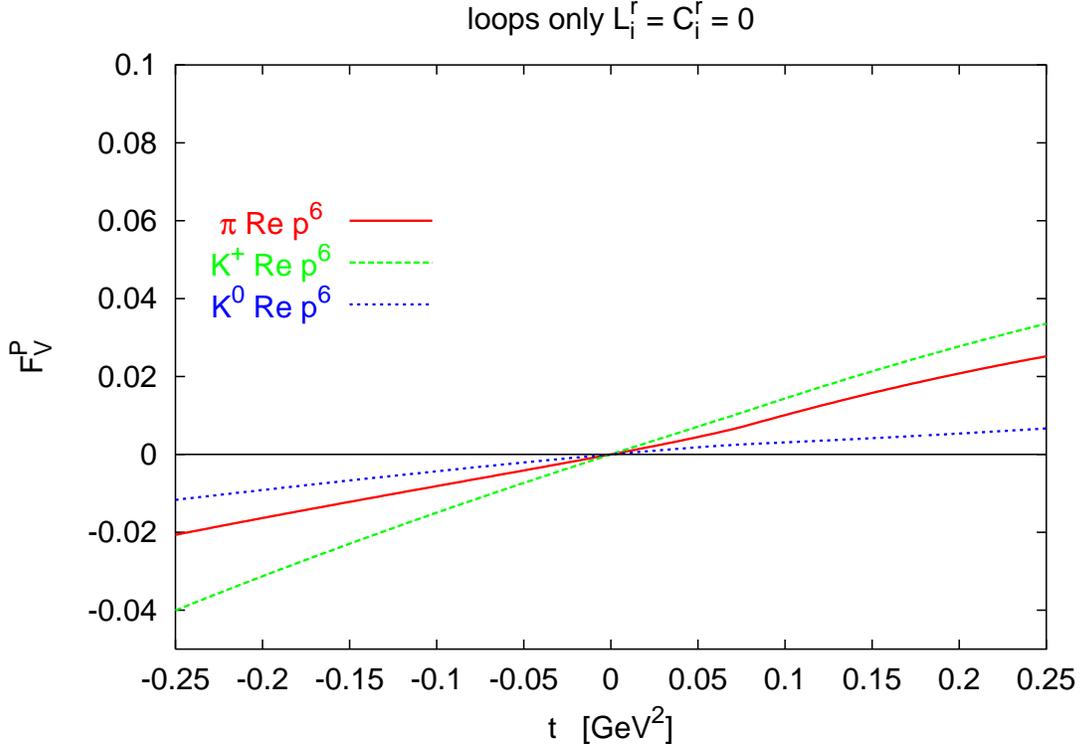}
\end{center}
\caption{The real parts of the loop diagrams
 at $\order(p^6)$ with all $L_i^r=0$,
for the pion, charged kaon and neutral kaon  form factor.
The scale is the same as in Fig.~\ref{figloops2}.
\label{figloops3}}
\end{figure}

We can now check how large the various loop contributions are.
In Fig. \ref{figloops1} we have shown the real and imaginary parts
of the pure loop contributions at $\order(p^4)$ and $\order(p^6)$
for the pion form factor.
To put the figures in perspective one should bear in mind that
the lowest order is exactly one and that at $t=\pm0.25$~GeV$^2$
the $\order(p^4)$ contribution from $L_9^r$
setting $L_9^r=6.9\times 10^{-3}$ is
$\pm0.40$. We show the difference between the contribution from
the pure loops only for the three different pseudoscalars in
Fig. \ref{figloops2} at $\order(p^4)$ and in Fig. \ref{figloops3}
at $\order(p^6)$.
\begin{figure}[t]
\begin{center}
\includegraphics[width=0.9\textwidth]{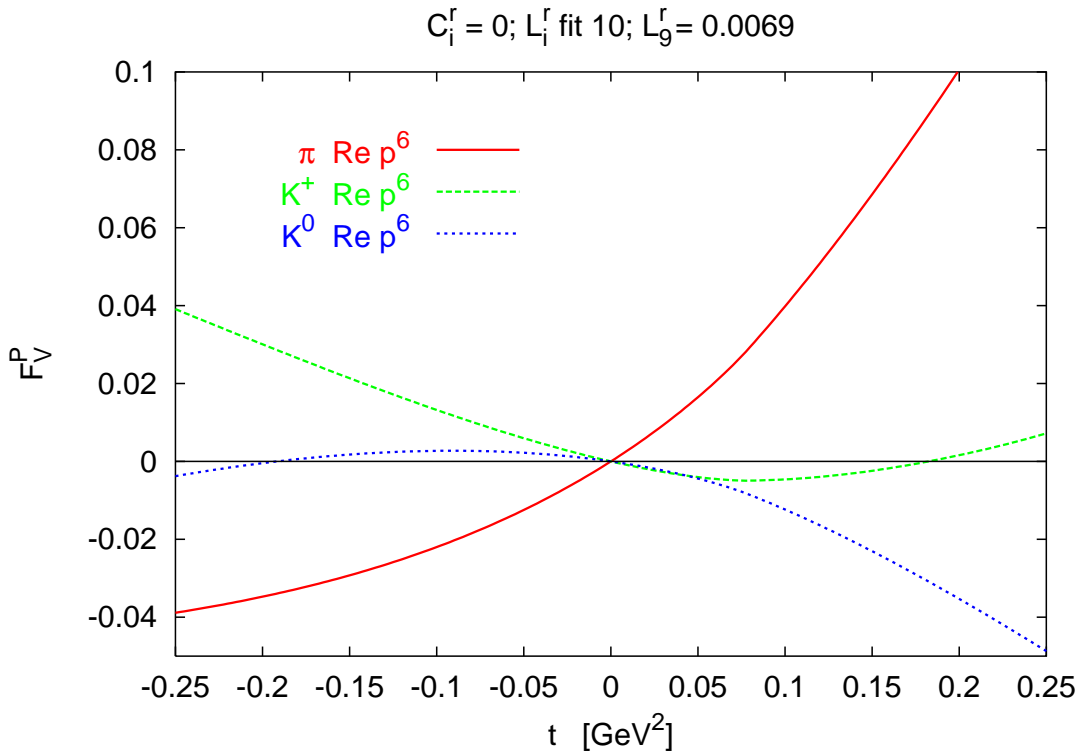}
\end{center}
\caption{The real parts
at $\order(p^6)$ 
for the pion, charged kaon and neutral kaon form factor
that depend on $L_i^r$. This corresponds to
$F_{V\mathbf{L}}^P(t)$. The inputs are taken from \cite{ABT4}.
The scale is the same as in Fig.~\ref{figloops2}.
\label{figloops4}}
\end{figure}
Fig. \ref{figloops4} shows the contribution from the $L_i^r$ dependent
part at $\order(p^6)$, $F_{V\mathbf{L}}^P(t)$ with the input parameters
as given in
\cite{ABT4}.

The part that depends on the ${\cal L}_6$ constants $C_i^r$
is also rather short.
\ba
\label{ctspp}
   F_{V\mathbf{C}}^\pi(q^2)& =&
 - 4 q^2 m_\pi^2 ( 4  C^r_{12} + 4 C^r_{13} + 2 
    C^r_{63} + C^r_{64} + C^r_{65} + 2 
    C^r_{90} )
 \nonumber\\ &&
 - 8 q^2 m_K^2 ( 4 C^r_{13} + C^r_{64} ) - 4
     q^4  ( C^r_{88} - C^r_{90} )\,,
\ea
\ba
\label{ctskp}
   F_{V\mathbf{C}}^{K^+}(q^2)& =&
 -  q^2 m_\pi^2 \left(  16 C^r_{13} +\frac{8}{3}  C^r_{63}
   + 4 C^r_{64} - \frac{4}{3} C^r_{65}\right)
 \nonumber\\ && 
- q^2 m_K^2 \left(16 C^r_{12} + 32 C^r_{13} + \frac{16}{3}C^r_{63} 
 +8 C^r_{64} +\frac{16}{3} C^r_{65} +8 C^r_{90} \right)
\nonumber\\&&
- 4     q^4  \left( C^r_{88} - C^r_{90} \right) \,,
\ea
and
\ba
\label{ctsko}
   F_{V\mathbf{C}}^{K^0}(q^2)& =&
   \frac{8}{3} q^2 (m_\pi^2-m_K^2) (  2 C^r_{63} - C^r_{65} )
 \,.
\ea

At low-energies, for instance the space-like interval with 
$\sqrt{-t} \lapprox 350$ MeV, the pion vector form factor
can be expanded in a Taylor series as
\begin{equation}
\label{radii}
F_V^\pi(t) = 1 + \frac{1}{6} \langle r^2 \rangle_V^\pi t 
+ c_V^\pi t^2 + \order(t^3)\,,
\end{equation}
defining in that manner the radii,$\langle r^2 \rangle_V^P$, 
and the coefficient $c_V^P$. 

\subsection{Isospin breaking}

Throughout this paper we work in the isospin limit.
The isospin breaking contributions can be split naively into two parts.

A first contribution comes through electromagnetic interactions.
This has been studied for the pion form factor in \cite{em}.
This
is primarily obtained by taking photon loops that develop an
infra-red singularity. Near threshold in the time-like region
this contribution can dominate over the next-to-next-to-leading terms
because of Coulomb final state interactions.
This is not the case 
for the pion form factor itself, the corrections are about $1\%$. For
the constant $c_V^\pi$ there might be a large contribution coming from
resolution dependent terms but they
cancel to a large extent for typical
values of the detector resolution.
The electromagnetic contribution to the masses we take into account
indirectly, we use the charged pion and charged kaon mass
in our numerical analysis.

The second source of isospin breaking is from the
mass difference $m_u - m_d$. 
The pion form factor receives corrections starting only at
 $\order\left((m_d-m_u)^2\right)$
\cite{ademollo}. This together with the constraint (\ref{gauge})
means that corrections only come from loop diagrams
and are thus expected to be small. In the kaon form factors
isospin breaking could be larger but given the present experimental
accuracy is not expected to be relevant.

\section{Data and fits with the ChPT expression}
\setcounter{equation}{0}
\label{datafits}

\subsection{Data on $F_V^\pi$}

Data on the pion form factor have been collected in both time-like,
$t>0$, and space-like, $t<0$, regions. We discuss both in turn.

\subsubsection{Space-like}

In the space-like region the measurements have been done
using $\pi e\to\pi e$ scattering.
There are two
main experiments \cite{Dally} at
FNAL and NA7 at CERN \cite{Amendolia}. 
Both experiments are
based on pions scattering off the electrons of a liquid hydrogen target 
and the data agree in the overlap region.  The last one has
significantly smaller
error bars and dominates clearly all the fits.
There are also somewhat older FNAL data from the same experiment using a lower
energy pion beam \cite{Dallyold}. These have larger errors and are
compatible with the newer data. References to older data can be found
in both \cite{Amendolia} and \cite{Dally}.
These data were analyzed in both papers with a pole-fit
\be
\label{polef}
F_V^\pi(t) = 
\frac{n}{\left(1-\frac{1}{6}\langle r^2\rangle_V^\pi t\right)^2}\,,
\ee
where n refers to the normalization uncertainty.
Ref. \cite{Amendolia} obtained
\begin{equation}
\label{ram1}
\langle r^2\rangle_V^\pi = (0.431\pm 0.010)~\mbox{fm}^2
\end{equation}
when they constrained $n$ and
\begin{equation}
\label{ram2}
\langle r^2\rangle_V^\pi = (0.427\pm 0.010)~\mbox{fm}^2
\end{equation}
with $n$ left free.
The equivalent results of \cite{Dally} are
\begin{equation}
\label{rdal1}
\langle r^2\rangle_V^\pi = (0.439\pm 0.030)~\mbox{fm}^2
\end{equation}
when they constrained $n$ and
\begin{equation}
\label{rdal2}
\langle r^2\rangle_V^\pi = (0.384\pm 0.088)~\mbox{fm}^2
\end{equation}
with $n$ left free.
\begin{table}
\begin{center}
\begin{tabular}{|c|cc|}
\hline
Data   & $n$   & $\langle r^2 \rangle_V^{\pi}$ [fm$^2$]\\
\hline
FNAL \cite{Dally}   &  $0.973\pm0.033$ & $0.382\pm0.078$\\
                         & $\equiv 1$    & $0.443\pm0.018$\\
NA7 \cite{Amendolia} &$0.990\pm0.004$ & $0.428\pm0.011$\\
                         & $\equiv 1$    & $0.452\pm0.006$\\
\hline
\end{tabular}
\end{center}
\caption{The parameters of a simple pole fit to the $\pi e$
scattering data. Note the difference with the normalization free.
\label{tablepidata}}
\end{table}
\begin{figure}[t]
\includegraphics[width=0.9\textwidth]{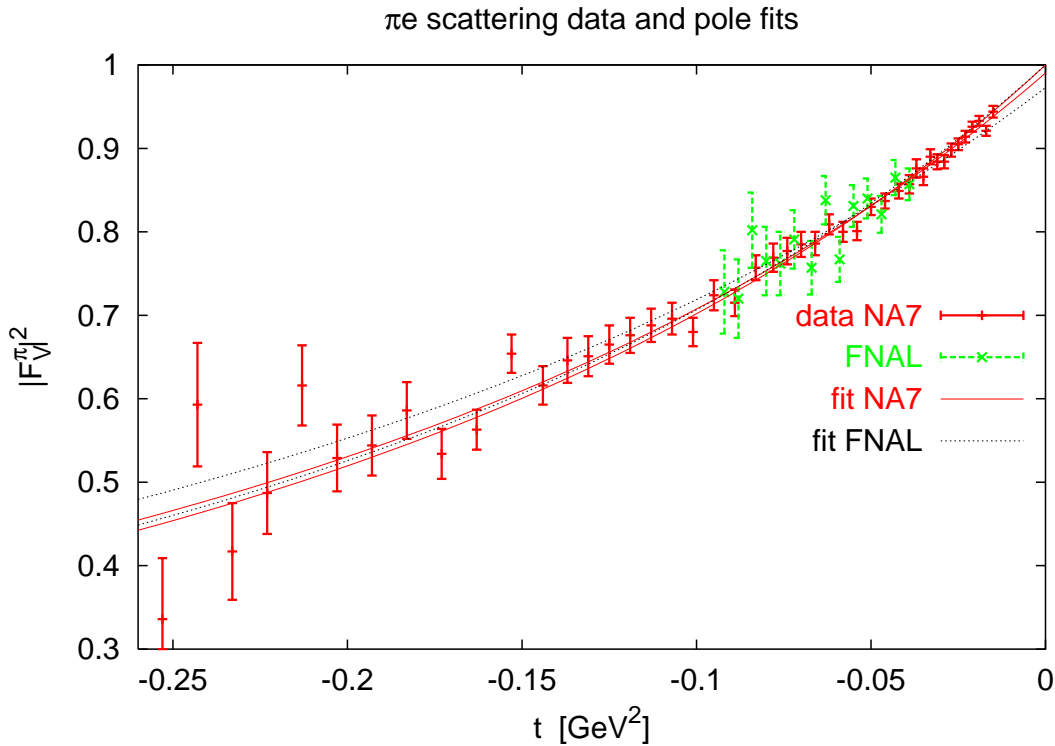}
\caption{The two data sets of $\pi e$ scattering and the resulting
pion form factor squared. Also shown are pole fits to both data sets.
\label{figpidata}}
\end{figure}
The two data sets together with the results of our own pole fits are
shown in Fig. \ref{figpidata} and the pole fit parameters are in Table
\ref{tablepidata}. The difference with the numbers quoted by the
experiments themselves is
because there the normalization was fixed using
the experimental error rather than set exactly equal to one.
The quoted errors for the normalization
are 1\% for \cite{Dally} and 0.9\% for \cite{Amendolia}.

Using the previous data sets there are several determinations of the
pion vector form factor in the
literature. The experiments were analyzed directly mainly with a pole
fit as discussed above. A full analysis within the framework
of two flavour ChPT is in Ref. \cite{BCT}, and we will compare with
the results from there in Sect. \ref{discussion}.

In \cite{Amendolia} the data 
were also analyzed using a dispersion relation technique with a
Pad\'e approximation for the Omn\'es representation
and $\pi\pi$ data
for the $\delta_1^1$ phase shift the result they obtained is
\begin{equation}
\label{ram3}
\langle r^2\rangle_V^\pi = (0.439\pm 0.008)\,\mbox{fm}^2\,.
\end{equation}
Details of this technique can be found in \cite{chez}.

\subsubsection{Time-like}

In the time-like region there are several experiments,
but unfortunately most of the data is located
outside the range of applicability of ChPT. Essentially the data
is obtained from two sources:
$\tau \rightarrow \pi \pi \nu_\tau$ \cite{tau}
and $e^- e^+ \rightarrow \pi^+\pi^-$ \cite{pi1,pi2,pi3,pi4}.
In the later experiment, \cite{pi4}, corresponding to the NA7 collaboration,
the systematic errors are close to the statistical ones and we believe
that they have been underestimated in the time-like region as they are
difficult to fit together with the space-like data.

A resonance based model 
was used in \cite{pi2} to fit the data up to 1.2 GeV
and determines the mass and decay constants
of the
$\rho$ as well as the pion radius with the result
\begin{equation}
\label{rvmd1}
\langle r^2\rangle_V^\pi = (0.460\pm 0.011)\, \mbox{fm}^2\,.
\end{equation}
Instead one can rely on
$\tau \rightarrow \pi \pi \nu_\tau$ data to fit the expression
\begin{equation}
F_V^\pi (t)= exp\left\{ \alpha_1 t +\frac{1}{2} \alpha_2 t^2 + \frac{t^3}{\pi}
\int_{4 m_\pi^2}^\infty \frac{dz}{z^3}
 \frac{\delta_1^1}{z-t-i\epsilon}\right\}\,,
\end{equation}
with $\alpha_1, \alpha_2$ given in terms of $\langle r^2\rangle_V^\pi$
and $c_V^\pi$ (see \cite{toni})
and $\delta_1^1$ determined by resonance saturation.
Keeping only the $\rho$ contribution one obtains
\begin{equation}
\label{rtoni}
\langle r^2\rangle_V^\pi = (0.429\pm 0.01)\, \mbox{fm}^2\,.
\end{equation}

\subsection{Data on $F_V^{K^+}$}

The charged kaon form factor has been measured directly via
kaon electron scattering method by the same two main
experiments that measured the pion form factor.
The older experiment is a FNAL experiment \cite{Dally_kaon}
and reported a rather low kaon charge radius
obtained by a fit to a pole form factor
\be
\langle r^2 \rangle_V^{K^+} = (0.28\pm0.05)~\mbox{fm}^2\,.
\ee
They also quoted a dipole fit where the normalization was
left free with the result
\be
\langle r^2 \rangle_V^{K^+} = (0.42\pm0.20)~\mbox{fm}^2\,.
\ee
In \cite{Dally} a 
simultaneous analysis of the pion and kaon form factors
was performed from the obtained data quoting the difference 
\be
\label{rKdiff1}
\langle r^2 \rangle_V^{K^+}-\langle r^2 \rangle_V^{\pi}
= (-0.16\pm0.06)~\mbox{fm}^2\,.
\ee
This difference is claimed to be largely free of systematic errors.

\begin{figure}[t]
\includegraphics[width=0.9\textwidth]{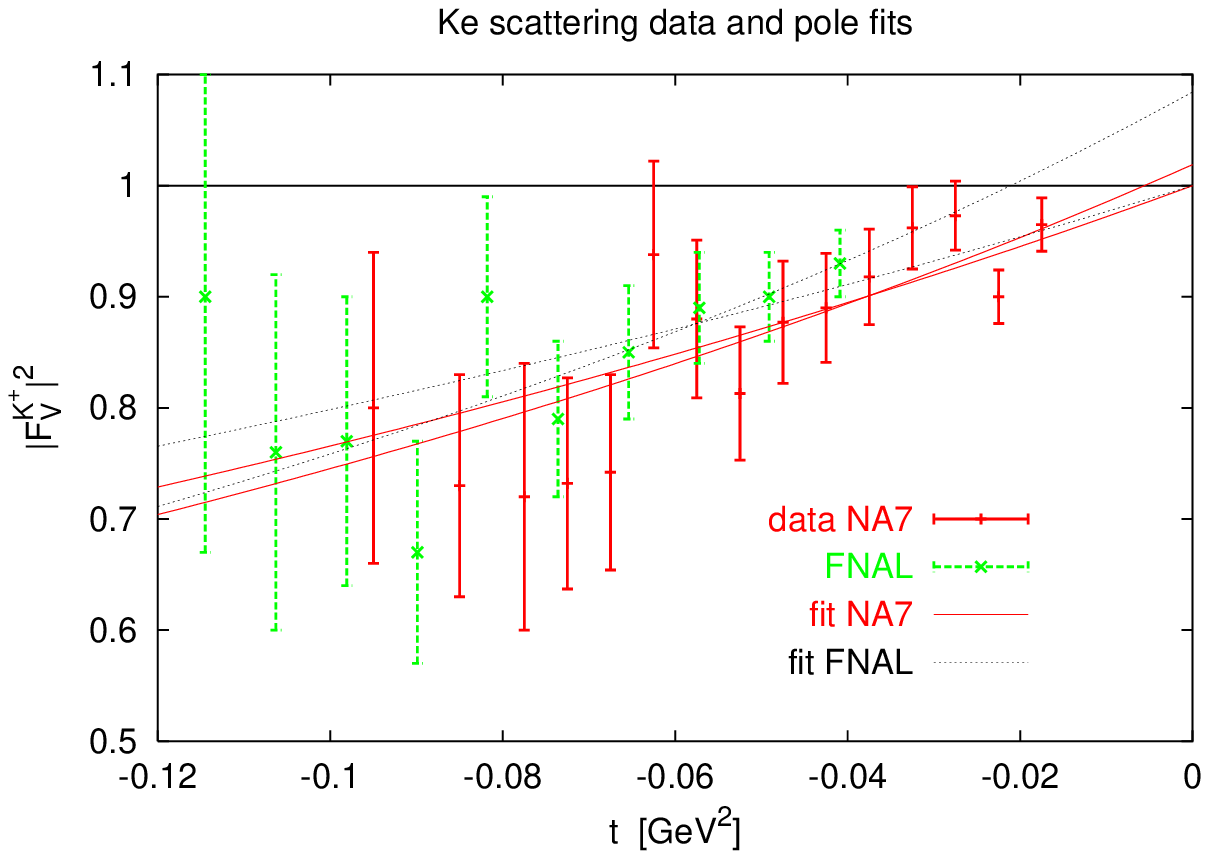}
\caption{The two data sets of $Ke$ scattering and the resulting
kaon form factor squared. Also shown are pole fits to both data sets.
\label{figKdata}}
\end{figure}

The more recent CERN experiment \cite{Amendolia_kaon}
finds using pole-like fits
\be
\langle r^2 \rangle_V^{K^+} = (0.34\pm0.05)~\mbox{fm}^2\,,
\ee
and from the ratio to the simultaneously collected pion data
\be
\label{rKdiff2}
\langle r^2 \rangle_V^{K^+}-\langle r^2 \rangle_V^{\pi}
= (-0.10\pm0.045)~\mbox{fm}^2\,.
\ee
Both data sets are shown in Fig. \ref{figKdata} together with a simple
pole fit with the normalization fixed to 1 and left free to both data sets.
The parameters of those fits are in Table \ref{tableKdata}.
\begin{table}
\begin{center}
\begin{tabular}{|c|cc|}
\hline
Data   & $n$   & $\langle r^2 \rangle_V^{K^+}$ [fm$^2$]\\
\hline
FNAL \cite{Dally_kaon}   & $1.084\pm0.091$ & $0.457\pm0.198$\\ 
                         & $\equiv 1$    & $0.278\pm0.046$\\
NA7 \cite{Amendolia_kaon} & $1.019\pm0.030$ & $0.395\pm0.109$\\
                          & $\equiv 1$    & $0.334\pm0.045$\\
\hline
\end{tabular}
\end{center}
\caption{The parameters of a simple pole fit to the $Ke$
scattering data. Note the difference with the normalization free.
\label{tableKdata}}
\end{table}

The threshold for kaon production in tau decays or $e^+e^-$ annihilation
is too high for the ChPT expressions to be valid. In principle a dispersion
relation could be used to relate those data with the threshold
parameters but this involves a rather large extrapolation. We have therefore
not included this type of data in our analysis.

\subsection{Data on $F_V^{K^0}$}

The data on the neutral kaon electromagnetic form factor are rather
scarce. There is a measurement of the radius using
kaon regeneration on electrons \cite{Molzon} leading
to
\be
\label{rK0}
\langle r^2 \rangle_V^{K^0} = (-0.054\pm0.026)~\mbox{fm}^2\,.
\ee
The decay $K_L\to\pi^+\pi^-e^+e^-$ also has contributions
that contain the neutral kaon electromagnetic form factor but
the extraction from the data is rather model dependent and not yet done
at present.

\subsection{Input parameters}
\label{input}

For the physical masses we use the charged pion and kaon mass \cite{pdg}
and the pion decay constant,
\be
m_{\pi^+} = 139.56995 ~\mbox{MeV}\,,
\quad
m_{K^+} = 493.677 ~\mbox{MeV}\,\mbox{ and }
F_{\pi} = 92.4 ~\mbox{MeV}\,.
\ee
For the eta mass we use the mass resulting from the GMO
relation. As discussed above, this does not make any difference within
the accuracy of our results.

A somewhat more complicated issue is how to deal with the input values
of the $L_i^r$ other than $L_9^r$.
We use the two-loop accuracy determination given
in fit 10 of \cite{ABT4}. 
We choose fit 10 and not the main fit, because the analysis
of the E865 experiment has been confirmed in the meantime
\cite{e865}\footnote{Note that we
only use the linear fit of the experimental slopes
and derivatives at threshold for the
form factors and not
their quadratic fit whose curvature can \emph{not} be reproduced by
the next-to-next-to-leading
calculation in ChPT of the $K_{\ell 4}$ form factors
with reasonable choices of the parameters.}.
Notice that in doing so, i.e. using the values in \cite{ABT4}
without any new refitting, we have
assumed that the value of $L_9^r$ does not have a substantial impact
on the rest of LECs.
This is true, as we discussed in \cite{ABT3}.
For the quantities studied in \cite{ABT4} $L_9^r$ 
only appears inside the $K_{\ell 4}$ form factors,
but together with the squared effective mass of the 
dilepton system and varying $L_9^r$ over an extremely wide range did not change
the fitting results.

\subsection{ChPT fits to $F_V^\pi(t)$}

Checking the $L_9^r$ dependence of $F_V^\pi$, we find that
the real part of this can be well described by a polynomial
of the form
\be
\label{L9polynomial}
\left. F_V^\pi(t)\right|_{p^6,\,L_9^r} = L_9^r\,
\left(a t +b t^2 +c t^3\right)\,.
\ee
The coefficients $a$, $b$ and $c$ are in principle dependent on the other
$L_i^r$ used as input.

We fit the data on $F_V^\pi(t)$ with the ChPT expressions with
the value of $L_9^r$ set equal to zero and a polynomial of the form
\be
\label{fs}
a_f t + b_f t^2 +c_f t^3
\ee
for several variations of the input. In this way we can separate the
dependence of $L_9^r$ on the unknown $\order(p^6)$ constants.
We fit similarly at $\order(p^4)$.

In the fit we have left the normalization of the data from
\cite{Amendolia}, \cite{Dally} and \cite{pi4} as an additional
fit parameter.

\begin{table}
\begin{center}
\begin{tabular}{|c|ccc|ccc|}
\hline
           &  $n$\cite{Amendolia} & $n$\cite{Dally} & $n$\cite{pi4}
                & $a_f$  [GeV]$^{-2}$& $b_f$ [GeV]$^{-4}$ 
              & $c_f$ [GeV]$^{-6}$ \\
\hline
$\order(p^4)$      & 
      0.981 & 0.979 & 1.073 & $1.285\pm0.003$ & $\equiv 0$ & $\equiv 0$\\ 
$\order(p^4)$      & 
      0.995 & 0.998 & 0.862 & $1.647\pm0.054$ & $3.83\pm0.28$ & $5.98\pm1.69$\\
$\order(p^6)$      &
      1.001 & 1.009 & 0.871 & $1.788\pm0.045$ & $3.16\pm0.27$ & $\equiv 0$\\
$\order(p^6)$      &
      0.994 & 0.998 & 0.860 & $1.693\pm0.054$ & $3.45\pm0.28$ & $5.33\pm1.69$\\
$\order(p^6)$ fit 11     &
      0.994 & 0.998 & 0.860 & $1.696\pm0.054$ & $3.44\pm0.28$ & $5.32\pm1.69$\\
$\order(p^6)$ fit 12     &
      0.994 & 0.998 & 0.860 & $1.704\pm0.055$ & $3.43\pm0.28$ & $5.30\pm1.72$\\
$\order(p^6)$ fit 13     &
      0.994 & 0.998 & 0.860 & $1.690\pm0.054$ & $3.45\pm0.28$ & $5.34\pm1.69$\\
polynomial only &
      0.995 & 0.998 & 0.871 & $1.936\pm0.056$ & $4.34\pm0.28$ & $6.44\pm1.69$\\
$\order(p^6)$ space &
      1.000 & 1.005 & --    & $1.844\pm0.018$ & $5.53\pm0.16$ & $12.3\pm1.0$\\
\hline
\end{tabular}
\end{center}
\caption{The results of fitting to $F_V^\pi(t)$ with different
theoretical inputs as explained in the text. $\equiv \mbox{a}$ stands for
a value held fixed equal to a.
\label{tabfitFpi}}
\end{table}

The fits we have performed are a pure $\order(p^4)$ fit, an $\order(p^4)$ fit
but allowing also
a $t^2$ and $t^3$ polynomial, and the same for the $\order(p^6)$ expressions.
For the input of the $L_i^r$ we used fit 10 of \cite{ABT4}.
The fits to the other $L_i^r$ with varying the assumptions
on $L_4^r$ and $L_6^r$ from that reference make essentially
no change as can be seen from the rows labelled fit 11, fit 12 and fit 13.
We also presented a pure polynomial fit. The $\chi^2$ for all the fits
including the $t^3$ term is the same and the quality of the fit
is good. The $\chi^2$ is 60.3 for 75 degrees of freedom.
The last line indicates the impact of the time-like data. This fit was
performed with the space-like data only and had a $\chi^2=48.6$
for 54 degrees of freedom.

Notice that the fits are somewhat different, but compatible within errors,
with the ones given in \cite{BCT}. The reason for this difference is that
the fits there were performed with the data of \cite{pi4} with
the normalization fixed to one. We have left that free.

\subsection{ChPT fits to $F_V^{K^+}$}

We now perform the same fits to the kaon data. The time-like data
are very much outside the domain of validity of ChPT so they are not
included. The space-like data are much less precise
than the pion ones and have a much smaller
range of $t$. There will thus be much less information about the
higher powers of $t$. For the pion we could fit using the whole region
$-0.25\le t \le0.25$, while here we have only the range $-0.11\le t\le-0.02$.
More care is needed in performing the fits, for instance, fitting with a $t^3$
term and both normalizations free leads to rather nonsensical results
as can be seen in the sixth row of Table \ref{tabfitFK}.

\begin{table}
\begin{center}
\begin{tabular}{|c|cc|ccc|}
\hline
           &  $n$\cite{Amendolia_kaon} & $n$\cite{Dally_kaon}  
                & $a_f$ [GeV]$^{-2}$ & $b_f$ [GeV]$^{-4}$ 
             & $c_f$ [GeV]$^{-6}$ \\
\hline
$\order(p^4)$      & 
      1.015 & 1.055 &  $1.408\pm0.300$ & $\equiv 0$ & $\equiv 0$\\ 
$\order(p^6)$      &	      
      0.993 & 1.028 &  $0.897\pm1.169$ & $-5.80\pm11.60$ & $\equiv0$\\
$\order(p^6)$      &	      
      1.015 & 1.055 &  $1.464\pm0.299$ & $\equiv0$ & $\equiv0$\\
$\order(p^6)$      &	      
      1.027 & 1.070 &  $1.756\pm0.297$ & $\equiv3$ & $\equiv0$\\
polynomial only &     
      0.890 & 0.919 &  $-3.259\pm3.989$ & $-96.9\pm82.0$ & $-559\pm492$\\
polynomial only &     
      0.993 & 1.028 &  $0.978\pm1.169$ & $-5.51\pm11.60$ & $\equiv0$\\
polynomial only &     
      1.014 & 1.054 &  $1.515\pm0.300$ & $\equiv0$ & $\equiv0$\\
\hline
\end{tabular}
\end{center}
\caption{The results of fitting to $F_V^{K^+}(t)$ with different
theoretical inputs as explained in the text.
\label{tabfitFK}}
\end{table}

\subsection{Results for the charge radii and $c_V^P$}

The NLO ChPT expression for the pion form factor \cite{GL2,BC}
was already analyzed quite
some time ago, using the space-like data \cite{BC}. The results were 
\begin{equation}§
\label{rp4old}
\langle r^2\rangle_V^\pi = 0.392\, (0.366)\, ~\mbox{fm}^2\,,
\end{equation}
with a fit with normalization one (free). 

The analysis using two flavour ChPT at two-loops led to the
conclusions \cite{BCT}
\be
\label{rtwoflavour}
\langle r^2\rangle_V^\pi = (0.437\pm0.016)~\mbox{fm}^2\,,
\quad
c_V^\pi =(3.85\pm0.60)~\mbox{GeV}^2
\ee
including
the data of \cite{pi4} and
\be
\label{rtwoflavour2}
\langle r^2\rangle_V^\pi = (0.453\pm0.016)~\mbox{fm}^2\,,
\quad
c_V^\pi =(4.45\pm0.60)~\mbox{GeV}^2
\ee
without including them.

The polynomial fit to all data gives
\be
\label{rpoly}
\langle r^2\rangle_V^\pi = (0.452\pm0.013)~\mbox{fm}^2\,,
\quad
c_V^\pi =(4.34\pm0.28)~\mbox{GeV}^2\,.
\ee
The pure $\order(p^4)$ fit yielded
\be
\label{rp4}
\langle r^2\rangle_V^\pi = (0.368\pm0.001)~\mbox{fm}^2\,,
\ee
while
the $\order(p^6)$ fit including the $t^3$ term leads to
\be
\label{rp6}
\langle r^2\rangle_V^\pi = (0.452\pm0.013)~\mbox{fm}^2\,,
\quad
c_V^\pi =(4.49\pm0.28)~\mbox{GeV}^2\,.
\ee
This last number we consider our final result for these quantities.
We have reproduced the previous fits with similar underlying assumptions.

The kaon charge radius obtained from the linear fit is
\be
\label{rKlin}
\langle r^2\rangle_V^{K^+} = (0.354\pm0.071)~\mbox{fm}^2\,,
\ee
while fitting the $\order(p^4)$ expression leads to
\be
\label{rKp4}
\langle r^2\rangle_V^{K^+} = (0.361\pm0.071)~\mbox{fm}^2\,,
\ee
and the $\order(p^6)$ fit to 
\be
\label{rKp6}
\langle r^2\rangle_V^{K^+} = (0.363\pm0.072)~\mbox{fm}^2\,.
\ee
Finally, 
fixing the quartic slope to a value similar to the one
obtained from the pion, we obtain
\be
\label{rKp62}
\langle r^2\rangle_V^{K^+} = (0.431\pm0.071)~\mbox{fm}^2\,.
\ee

\section{Discussion and Determination of $L_9^r$}
\label{discussion}
\setcounter{equation}{0}

The ChPT expression for the form factors depends on the unknown
constants $C_i^r$. For the remainder of the discussion we define
\ba
F_{V\mathbf{C}}^\pi(q^t)& =& R_{V1}^\pi q^2 + R_{V2}^\pi q^4\,,
\nonumber\\
F_{V\mathbf{C}}^{K^+}(q^t)& =& R_{V1}^{K^+} q^2 + R_{V2}^{K^+} q^4\,,
\nonumber\\
F_{V\mathbf{C}}^{K^0}(q^t)& =& R_{V1}^{K^0} q^2 + R_{V2}^{K^0} q^4\,,
\ea
and the only model-independent relations are
\be
R_{V2}^{\pi} = R_{V2}^{K^+}\quad\mbox{and}~R_{V2}^{K^0}=0\,.
\ee

\subsection{Estimates of the $R_{Vi}^P$}
\label{estimates}
\label{resonance}

In this section we estimate \emph{some} $\order(p^6)$ constants.
In principle these constants should be obtained
from QCD directly but this is not possible at present.
This, together with the
large number of LECs at higher orders in the chiral expansion,
 makes it unavoidable to
resort to model estimates to gain some predictability.
We stress
that this model 
dependence only starts at $\order(p^6)$. At $\order(p^4)$
we use the data directly.
The main idea is to saturate the $\order(p^6)$ order parameters via the
exchange of higher
mass resonances \cite{reso,reso2}. 
We use
the notation from \cite{reso} to which we refer for a more
extensive exposition of the method. 
One uses a matter field Lagrangian coupled to the Goldstone Bosons octet
\be
{\cal L}_{\mbox{\tiny{res}}} = 
\sum_{V,A,S,P} \left\{ {\cal L}_{\mbox{\tiny{Kin}}}(R)
+{\cal L}_{\mbox{\tiny{Int}}}\right\}\,,
\ee
where for the cases we are interested in, the allowed intermediate states are
 reduced to vectors. 
For these spin-1
mesons we use the realization where the vector contribution to the chiral
Lagrangian starts at $\order(p^6)$. We shall specifically only discuss terms
relevant
for our purposes.
\ba
{\cal L}_V &=& -\frac{1}{4}\langle V_{\mu\nu}V^{\mu\nu}\rangle
+\frac{1}{2}m_V^2\langle V_\mu V^\mu\rangle
-\frac{f_V}{2\sqrt{2}}\langle V_{\mu\nu}f_+^{\mu\nu}\rangle
\nonumber\\
&&-\frac{ig_V}{2\sqrt{2}}\langle V_{\mu\nu}[u^\mu,u^\nu]\rangle
+f_\chi\langle V_\mu[u^\mu,\chi_-]\rangle\,,
\label{vector}
\ea
with
\ba
V_{\mu \nu}  =  \nabla_\mu V_\nu - \nabla_\nu V_\mu \,,& &
f^{\mu \nu}_{\pm}  =
u (v^{\mu \nu}-a^{\mu \nu})  u^\dag
\pm u^\dag (v^{\mu \nu}+a^{\mu \nu}) u \nonumber \,,
\ea
\noindent
and
$V_\mu$ is a three-by-three
matrix describing the full vector nonet.
Furthermore $m_V$ refers to the octet and singlet
masses in the chiral limit.
The matrix $u_\mu$ is defined in terms of the covariant derivative,
Eq.~(\ref{covariant}),
\be
u_\mu = i u^\dagger D_\mu U u^\dagger = u_\mu^\dagger\,,\quad u^2 = U\,.
\ee

Instead of evaluating each contribution diagrammatically we
integrate out the heavy degree of freedom by means
of the classical equation of motion obtaining
\be
{\cal L}_V =
- \frac{i g_V f_V}{2 m^2_V} \langle ( \nabla_\lambda f_+^{\lambda \nu} )
(\nabla^\mu [ u_\mu, u_\nu ] ) \rangle
- \frac{f_\chi f_V}{\sqrt{2} m^2_V}
\langle ( \nabla_\lambda f_+^{\lambda \mu} )
[ u_\mu, \chi_- ] \rangle \; ,
\label{LagInt}
\ee
where the resulting Lagrangian is directly of $\order(p^6)$.

The numerical values for the couplings constants appearing
in (\ref{LagInt}) are obtained using
experimental information. The value
of $f_V$ is obtained from
$\rho \rightarrow e^+ e^-$ \cite{reso2}. The parameters $f_\chi$ and $g_V$
are determined
simultaneously from $\rho \rightarrow \pi \pi$ and
$K^* \rightarrow K \pi$ \cite{BCEGS2}. 
In summary we use
\begin{eqnarray}
f_V = 0.20\,,\quad  f_\chi = -0.025\,,\quad  g_V = 0.09\,,
\end{eqnarray}
and the vector-mass is taken as the experimental one \cite{pdg}.
\begin{eqnarray}
m_V = m_\rho = 0.77 \mbox{ GeV}\,.
\end{eqnarray}
The description is basically identical to the one in \cite{BCT} but
now also includes the kaon case.
Numerically we obtain
\ba
\label{RVi}
R_{V2}^{\pi} &=& R_{V2}^{K^+} = 0.26\times 10^{-3}\,,
\nonumber\\
R_{V1}^{\pi} &=& -0.49\times 10^{-5}~\mbox{GeV}^2\,,
\nonumber\\
R_{V1}^{K^+} &=& -6\times 10^{-5}~\mbox{GeV}^2\,,
\nonumber\\
R_{V1}^{K^0} &=& 0  \,.
\ea

An alternative estimate is based on the full VMD form for the form factors
\ba
\label{FVVMD}
F_V^\pi(t) &=& \frac{m_\rho^2}{m_\rho^2-t}
\nonumber\\
F_V^{K^+}(t) &=& \frac{1}{2}\frac{m_\rho^2}{m_\rho^2-t}
+\frac{1}{6}\frac{m_\omega^2}{m_\omega^2-t}
+\frac{1}{3}\frac{m_\phi^2}{m_\phi^2-t}\,,
\nonumber\\
F_V^{K^0}(t) &=& -\frac{1}{2}\frac{m_\rho^2}{m_\rho^2-t}
+\frac{1}{6}\frac{m_\omega^2}{m_\omega^2-t}
+\frac{1}{3}\frac{m_\phi^2}{m_\phi^2-t}\,.
\ea
With $m_V$ as above the chiral limit of the vector meson masses,
this leads to
\ba
\label{RViVMD}
R_{V2}^{\pi} &=& R_{V2}^{K^+} = \frac{F_\pi^4}{m_V^4} 
\approx 0.21\times 10^{-3}\,,
\nonumber\\
R_{V1}^{\pi} &=& - \frac{F_\pi^4}{m_V^4}\left(m_\rho^2-m_V^2\right)
\approx -0.35\times 10^{-5}~\mbox{GeV}^2\,,
\nonumber\\
R_{V1}^{K^+} &=& - \frac{F_\pi^4}{m_V^4}
\left(\frac{1}{2}m_\rho^2
+\frac{1}{6}m_\omega^2+\frac{1}{3}m_\phi^2-m_V^2\right)
\approx  -3\times 10^{-5}~\mbox{GeV}^2\,,
\nonumber\\
R_{V1}^{K^0} &=&  - \frac{F_\pi^4}{m_V^4}
\left(-\frac{1}{2}m_\rho^2
+\frac{1}{6}m_\omega^2+\frac{1}{3}m_\phi^2-m_V^2\right)
\approx -3\times 10^{-5}~\mbox{GeV}^2\,.
\ea
We used here $m_\rho^2-m_V^2\approx (m_{K^*}^2-m_\rho^2)m_\pi^2/m_K^2$.
The pion numbers of (\ref{RViVMD}) and (\ref{RVi}) are in reasonable agreement.
The numbers for the kaons have a sizable uncertainty. 
E.g. changing $m_V$ to $m_\phi$ in the denominator changes the values
by
a factor of $\left(m_\rho/m_\phi\right)^4 \approx 0.3$, which
differs from one
only by higher order effects. We will use the values for $R_{V1}^\pi$ and
$R_{V2}^\pi$ but not for the kaons, sizable higher order quark mass
effects are known to exist in the vector meson sector \cite{CHPTVector}.

\subsection{Obtaining $L_9^r$}

At $\order(p^4)$ the relation between $a_f$, Eq. (\ref{fs}), and $L_9^r$
is direct, $2L_9^r/F_\pi^2 = a_f$ and we obtain
\be
\label{L9p4}
\left.L_9^r(0.77~\mbox{GeV})\right|_{p^4\mbox{\small fit}} =
5.5~(7.0)\times 10^{-3}\,. 
\ee
The number in brackets is from the fit with $b_f$ and $c_f$ left free.
The value obtained from the kaon form factor is perfectly compatible
with this since the values of $a_f$ are compatible within errors.

At the next-to-next-to-leading order the relation is a little different
\be
a_f = \frac{2}{F_\pi^2} \left(1+\Delta_9\right) L_9^r + 
\frac{1}{F_\pi^4}R_{V1}^\pi\,.
\ee
The value of $\Delta_9$ follows from our calculation and depends explicitly
on the values of the
other $L_i^r$. For the inputs of fit 10 in \cite{ABT4} we obtain
\be
\Delta_9 = 0.27\,.
\ee
This together with the value of $R_{V1}^\pi$ of (\ref{RVi}) leads
from the $\order(p^6)$ fitted value of $a_f$ to
\be
\label{L9p6}
\left.L_9^r(0.77~\mbox{GeV})\right|_{p^6\mbox{\small fit}} =
6.25~(5.93)\times 10^{-3}\,, 
\ee
where the number in brackets is the fit including the $t^3$ term.
The difference between the two numbers is an indication of the $\order(p^8)$
and higher corrections. Neglecting $R_{V1}^\pi$ would have lowered
both these numbers by
\be
\left.L_9^r(0.77~\mbox{GeV})\right|_{R_{V1}^\pi} = 0.23\times 10^{-3}\,.
\ee
Taking the difference in the two values of $L_9^r$ in (\ref{L9p6})
as an estimate of the error from higher orders, allowing for
a factor of two uncertainty in $R_{V1}^\pi$ and adding the experimental
error, all in quadrature  leads
to
\be
\label{L9final}
L_9^r(0.77~\mbox{GeV}) = (5.93\pm0.43)\times 10^{-3}
\ee
as our final result. Notice that the error due to experiment only
is about half this.

Using this value for $L_9^r$ and the value of $b_f$ 
we can extract a measured value of $R_{V2}^\pi$ of
\be
R_{V2}^\pi = -4\left(C_{88}^r-C_{90}^r\right) =(0.22\pm0.02)\times10^{-3}\,,
\ee
in good agreement with the estimates (\ref{RVi}) and (\ref{RViVMD}).

The kaon data do not lead to any better determination of these parameters.
The extra constraints we obtain from them are given in
Section \ref{chptradii}.

\subsection{Charge radii and the $c_V^P$ in ChPT}
\label{chptradii}

We now present the different contributions to the charge radii.
{}For the pion charge radius the different contributions are
given by
\ba
\langle r^2\rangle_V^{\pi} &=& \Big\{
0.325[L_9~p^4]+0.068 [\mbox{loops }p^4]
+0.084 [L_9 ~p^6]
\nonumber\\
&&-0.011 [\mbox{loops }p^6]
-0.015 [R_{V1}^\pi]\Big\}~\mbox{fm}^2
\nonumber\\
&=& 0.451~\mbox{fm}^2\,,
\ea
where the terms inside square brackets reveal the source.
The charged kaon charge radius gives similarly
\ba
\langle r^2\rangle_V^{K^+} &=& \Big\{
0.325[L_9~p^4]+0.031 [\mbox{loops }p^4]
+0.018 [L_9 ~p^6]
\nonumber\\
&&-0.011 [\mbox{loops }p^6]
-0 [R_{V1}^{K^+}]\Big\}~\mbox{fm}^2
\nonumber\\
&=& (0.363\pm 0.19)~\mbox{fm}^2\,.
\ea
The error is given by using the value of (\ref{RVi}) for
$R_{V1}^{K^+}$.
We can in fact use these results to obtain an experimental
bound on the combination of $\order(p^6)$ constants containing terms 
with masses
\be
\langle r^2\rangle_V^{\pi}-\langle r^2\rangle_V^{K^+}
= 0.102~\mbox{fm}^2+\frac{6}{F_\pi^4}\left(R_{V1}^\pi-R_{V1}^{K^+}\right)\,.
\ee
With the experimental value (\ref{rKdiff2}) we obtain
\be
R_{V1}^\pi-R_{V1}^{K^+} = \left(0.0\pm1.4\right)\times 10^{-5}~\mbox{GeV}^2
\ee
or with the experimental value (\ref{rKdiff1}) 
\be
R_{V1}^\pi-R_{V1}^{K^+} = \left(1.9\pm1.9\right)\times 10^{-5}~\mbox{GeV}^2\,.
\ee
Both values are within a factor of two to three
of the estimates
done above in (\ref{RVi}) and (\ref{RViVMD}).

Our calculation can be used to predict the neutral kaon charge radius.
The $L_9^r$ dependence is basically zero at $\order(p^6)$ as well.
We thus obtain
\ba
\label{rpredK0}
\langle r^2\rangle_V^{K^0} &=&\Big\{ -0.0365[\mbox{loops } p^4]
-0.0057[\mbox{loops } p^6] \Big\} ~\mbox{fm}^2
+\frac{6}{F_\pi^4} R_{V2}^{K^0} 
\nonumber\\
&=& (-0.042\pm 0.012) ~\mbox{fm}^2\,.
\ea
The error is based on {\em assuming} that the unknown contribution
is not larger than twice the $\order(p^6)$ loop contribution.
Using the estimate for $R_{V1}^{K^0}$ of (\ref{RViVMD})
lowers the value of (\ref{rpredK0}) by about 0.10.
The result is in good agreement with the measurement (\ref{rK0}). 
We can also  turn the argument around and obtain
\be
R_{V2}^{K^0} = (-0.4\pm0.9)\times 10^{-5}~{GeV}^2\,,
\ee
which is compatible with the estimate (\ref{RVi}) and somewhat lower than
(\ref{RViVMD}).

The contribution to $c_V^P$ can be expanded similarly.
For the pion we obtain
\ba
c_V^\pi &=& \Big\{
0.66 [\mbox{loops }p^4]
+0.41 [L_9 ~p^6]
\nonumber\\
&&+0.38 [\mbox{loops }p^6]
+3.04 [R_{V2}^\pi]\Big\}~\mbox{GeV}^{-4}
\nonumber\\
&=&4.49~\mbox{GeV}^{-4}\,.
\ea
Here we used the estimate
for $R_{V2}^\pi$ obtained from the measurement (fifth row $b_f=3.45$) in Table
\ref{tabfitFpi} and removing the contribution from $L_9^r$.
For the charged kaon
\ba
c_V^{K^+} &=& \Big\{
0.37 [\mbox{loops }p^4]
+0.19 [L_9 ~p^6]
\nonumber\\
&&+0.11 [\mbox{loops }p^6]
+3.04 [R_{V2}^{K^+}]\Big\}~\mbox{GeV}^{-4}
\nonumber\\
&=&3.71~\mbox{GeV}^{-4}\,.
\ea
Note that the data as discussed are not good enough to give
a reasonable value for this parameter.
Finally we obtain for the neutral kaon
\ba
c_V^{K^0} &=& \Big\{
-0.29 [\mbox{loops }p^4]
-0.22 [L_9 ~p^6]
\nonumber\\
&&-0.23 [\mbox{loops }p^6]
+0.00 [R_{V2}^{K^+}]\Big\}~\mbox{GeV}^{-4}
\nonumber\\
&=&-0.74~\mbox{GeV}^{-4}\,.
\ea

\begin{figure}[t]
\begin{center}
\includegraphics[width=0.9\textwidth]{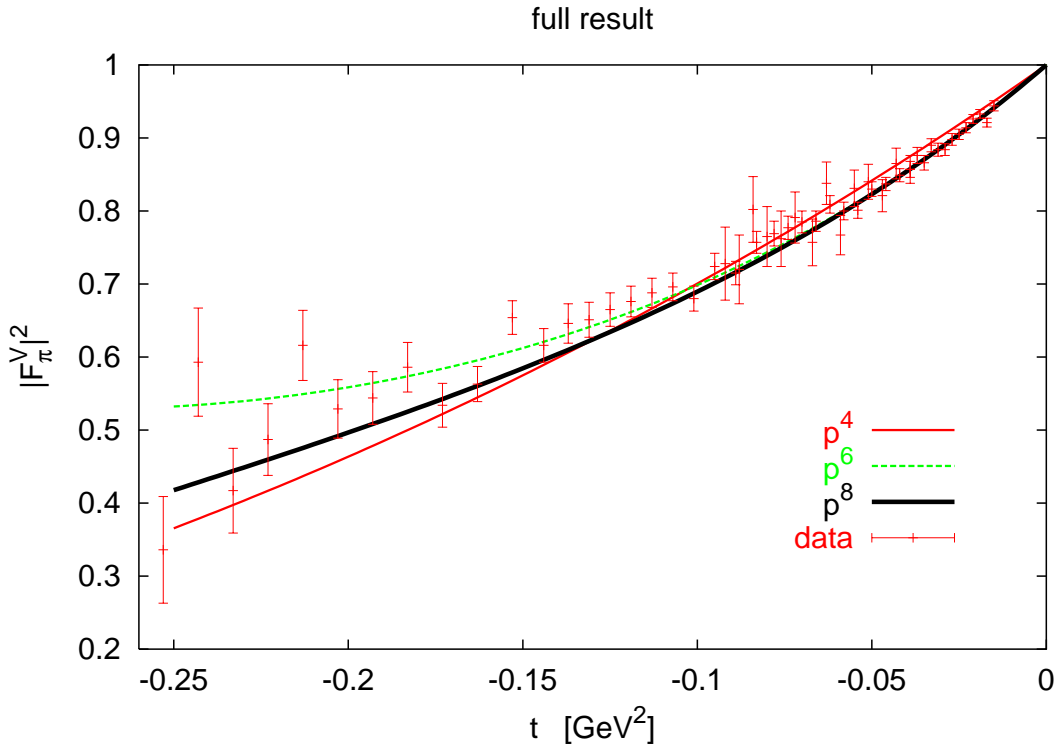}
\end{center}
\caption{\label{figspace} The comparison of the space-like measurements
with the ChPT calculation. Notice that there is excellent convergence
over the whole kinematical range.}
\end{figure}

\begin{figure}[t]
\begin{center}
\includegraphics[width=0.9\textwidth]{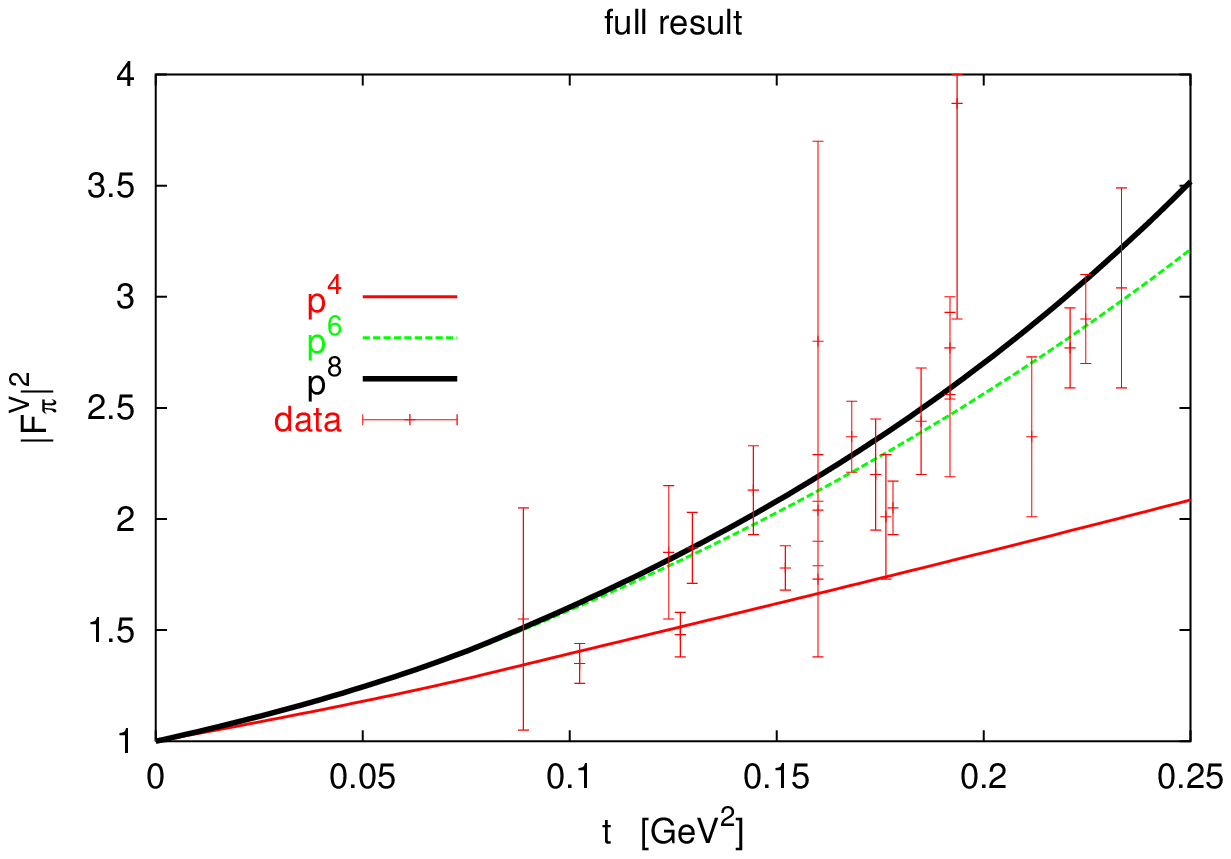}
\end{center}
\caption{\label{figtime} The comparison of the time-like measurements
with the ChPT calculation. Notice that there is excellent convergence
over the whole kinematical range and the slow buildup of the
tail of the $\rho$-meson. The data have not been corrected for
the normalization.}
\end{figure}

We also show the overall agreement with the data in Figs. \ref{figspace} and 
\ref{figtime} for the measured pion form factor. Note that we have plotted
the data with normalization one, not the fitted normalization.
As can be seen the convergence is nice and of a similar quality
as the two flavour results in \cite{BCT}. As stated before,
we are in excellent agreement with that reference when the difference
in the treatment of the data of \cite{pi4} is taken into account.

\subsection{Comparison with predictions for $L_9^r$}
\label{comparison}

Our estimates for the 
$\order(p^6)$ are based on a resonance saturation model.
As seen above, unless the estimates are off by more than a factor
of two, the value obtained for $L_9^r$ is reliable.
In table \ref{tablereso}, we compare the result (\ref{L9p6}) with several
model approaches.
First of all with the resonance saturation prediction \cite{reso}
that in this
particular case reduces to the Vector contribution 
\begin{equation}
L_9^r(m_V) = \frac{F_V G_V}{2 m_V^2}\,. 
\end{equation}
In addition if one implements 
some extra ``QCD inspired assumptions'' for the high energy behaviour,
such as an unsubtracted
dispersion relation for the pion form factor, one obtains \cite{reso2}
\begin{equation}
L_9^r(m_V) = \frac{F_\pi^2}{2 m_V^2}\,.
\end{equation}
We also show a constituent-quark model \cite{domenech} which in the
chiral and large-N$_c$ limit and including
leading gluon contributions leads to
\begin{equation}
L_9^r = \frac{N_c}{48\pi^2}[1+\order(1/M_Q^6)]\,,
\end{equation}
with $M_Q$ being a constituent quark mass.
Using only the first resonance plus a continuum spectrum to mimic QCD
one obtains the so call LMD model \cite{PPR}.
If one implements in addition QCD sum rules, one gets \cite{GP}
\begin{equation}
L_9^r(m_\rho) = \frac{1}{2} \frac{F_\pi^2}{m_\rho^2} =
 \frac{5}{2\sqrt{6}} \frac{1}{16\pi^2}\,.
\end{equation}
An extensive discussion can be found in the more
recent study of \cite{KN}.
At the expected accuracy all of these predictions are
in good agreement with the value obtained
in (\ref{L9final}).

\begin{table}[t]
\begin{center}
\begin{tabular}{|cccc|}
\hline
Eq.~(\ref{L9p6})&
\cite{reso}& 
\cite{PPR}, 
\cite{GP} &
\cite{domenech}
 \\
\hline
$5.93$ & $\sim7$ &  $6.5$ & $6.33$\\
\hline
\end{tabular}
\end{center}           
\caption{\label{tablereso} Values of $10^{-3}\times L_9^r(m_\rho)$
for different model
approaches in comparison with its value obtained with an $\order(p^6)$ fit.}
\end{table}

\section{Summary}
\setcounter{equation}{0}
\label{summary}

Let us summarize our findings. We have described
the evaluation at $\order(p^6)$
in ChPT of the pion and kaon electromagnetic form factor, giving an overview
of the methods and checks, both numerical and analytical.
The pion vector form factor
expression is quite handleable and we
have listed it completely, partly in  App.~\ref{appendixpion}
and the rest in Sect. \ref{nnlresults}.

We have assumed a resonance saturation model
to determine the Born contribution appearing at
$\order(p^6)$ and compared it with the $\order(p^6)$ parameters 
we could determine directly.  
There is a priori no reason to doubt that this or similar methods
retains the bulk of the $\order(p^6)$ counter terms. This point was
certainly fulfilled for
all the studies inside the two-flavour case.
What is evident is that the use of an effective
approach enforces to use an estimate for the higher
order parameters.
This constitutes at present the main source of theoretical uncertainty.

If we start with the premise of a ``well behaved'' series
expansion such terms are sub-sub-leading
and therefore the impact of their precise value ought to be mild.
Thus, to make some sense of the full approach,
there should be no need to fine tune
many of the $\order(p^6)$ constants. In this work, the impact of these
higher order parameters seemed reasonable.

For the pion form factor we have collected all the available
data with reasonable precision and
combined it with our analytical results to obtain
a new determination
of the pion charge radius and $c_V^\pi$,
as well as the LEC $L_9^r(m_\rho)$, Eq.~(\ref{L9p6}). 
All these results are compatible with previous ones.
In addition we have evaluated the correction due to higher order terms.
The genuine loop contributions do not show any strong deviation and 
they define a proper convergent expansion, see Fig.~\ref{figloops1}.

The data on the kaon form factor are also included and are within
errors perfectly well described by ChPT. We have used these
data to put {\em experimental}
bounds on two combinations of $\order(p^6)$ constants involving quark masses.

It is also evident from our discussion that the direct role 
of the Zweig rule suppressed terms with $L_4^r$ are marginal for the present
processes
because they are accompanied 
by products of $m_\pi^2$. The indirect impact via its influence on the
determination of the other $L_i^r$ \cite{ABT4} is also small here.

\section*{Acknowledgements}

We thank G.~Amor\'os for collaboration in the early stages of this project. 
This work was
supported in part by TMR, EC--Contract No.
ERBFMRX--CT980169 (EURODAPHNE).

\appendix
\renewcommand{\theequation}{\Alph{section}.\arabic{equation}}

\section{Pion vector form factor}
\label{appendixpion}
\setcounter{equation}{0}

In this appendix we give the explicit expressions for the
${\cal O}(p^6)$ parts of the pion form factor which were
not given in Sect. \ref{nnlresults}.

\subsection{Reducible contributions:\\
$\mbox{1-loop contribution} \times \mbox{1-loop contribution}$}

This contribution can be obtained
from diagrams (a,c,e,g,i) and (j) in Fig.~\ref{fig2}.
In terms of the integrals defined in App.~\ref{appintegrals} we obtain
\begin{eqnarray}
   F_{V\mathbf{B}}^\pi(q^2) &=&
        1/(16\pi^2)   \left\{  - 1/2\, m_\pi^2 \overline{A}(m_\pi^2) - 1/4\, m_\pi^2 \overline{A}(m_K^2) \right\}
\nonumber\\&&
       + 1/(16\pi^2)^2   \left\{  - 1/48\, m_\pi^2 (3m_K^2+10m_\pi^2) \pi^2 - 35/96\, m_\pi^2 (2m_K^2-m_\pi^2)  
\right.
\nonumber\\&&
\left.
         - 89/48\, m_\pi^4 
         - 1/16\, q^2 
         (m_K^2+2 m_\pi^2) (1 +\pi^2/6)\right\}
\nonumber\\&&
       + 1/(16\pi^2) \overline{B}_{22}^\epsilon(m_\pi^2,m_\pi^2,q^2) ( 5 m_\pi^2 - 1/2\, q^2 )
\nonumber\\&&
       + 1/(16\pi^2) \overline{B}_{22}^\epsilon(m_K^2,m_K^2,q^2) ( 3/2\, m_\pi^2 - 1/4\, q^2 )
\nonumber\\&&
       + 4 \overline{B}_{22}(m_\pi^2,m_\pi^2,q^2)^2   
%
       + 4 \overline{B}_{22}(m_\pi^2,m_\pi^2,q^2) \overline{B_{22}}(m_K^2,m_K^2,q^2) 
\nonumber\\&&
       + \overline{B}_{22}(m_\pi^2,m_\pi^2,q^2)\left\{  - 4 \overline{A}(m_\pi^2) - 2 \overline{A}(m_K^2) \right\}
%
       + \overline{B}_{22}(m_K^2,m_K^2,q^2)^2   
\nonumber\\&&
       + \overline{B}_{22}(m_K^2,m_K^2,q^2)  \left\{ - 2 \overline{A}(m_\pi^2) - \overline{A}(m_K^2) \right\}
\nonumber\\&&
       - 1/4\, \overline{A}(m_\pi^2)^2 - 3/8\, m_\pi^2/m_K^2 \overline{A}(m_K^2)^2 
        + \overline{A}(m_\pi^2) \overline{A}(m_K^2)  
\nonumber\\&&
          + 1/4\, \overline{A}(m_K^2)^2 
        - 1/8\, q^2/m_\pi^2 \overline{A}(m_\pi^2)^2 - 1/16\, q^2/m_K^2
          \overline{A}(m_K^2)^2\,.
\end{eqnarray}

\subsection{Irreducible contributions}

We have disentangled the contribution to the irreducible diagrams in
two pieces. The first one comes
from the topology (a) shown in Fig.~\ref{fig3}. The treatment of the
possible tensor integrals was
treated lengthly in \cite{ABT1} where we refer the reader for the notation.
This term reads
\begin{eqnarray}
   F_{V\mathbf{H}}^\pi =&&
        H^{F^\prime}(m_\pi^2,m_\pi^2,m_\pi^2;m_\pi^2)   ( 3/2\, m_\pi^4 )
\nonumber\\&&
       + H^{F^\prime}(m_\pi^2,m_K^2,m_K^2;m_\pi^2)   (  - 5/8\, m_\pi^4 )
\nonumber\\&&
       + H^{F^\prime}(m_\pi^2,m_\eta^2,m_\eta^2;m_\pi^2)   ( 1/18\, m_\pi^4 )
\nonumber\\&&
       + H^{F^\prime}(m_K^2,m_\pi^2,m_K^2;m_\pi^2)   ( m_\pi^2 m_K^2 )
\nonumber\\&&
       + H^{F^\prime}(m_K^2,m_K^2,m_\eta^2;m_\pi^2)   (  - 5/6\, m_\pi^4 )
\nonumber\\&&
       + H^{F^\prime}(m_\eta^2,m_K^2,m_K^2;m_\pi^2)   ( 1/4\, m_\pi^2 (2 m_K^2 - 1/2\, m_\pi^2) )
\nonumber\\&&
       + H^{F^\prime}_{1}(m_\pi^2,m_\pi^2,m_\pi^2;m_\pi^2)   (  - 2 m_\pi^4 )
\nonumber\\&&
       + H^{F^\prime}_{1}(m_\pi^2,m_K^2,m_K^2;m_\pi^2)   ( m_\pi^4 )
\nonumber\\&&
       + H^{F^\prime}_{1}(m_K^2,m_K^2,m_\eta^2;m_\pi^2)   ( 2 m_\pi^4 )
\nonumber\\&&
       + H^{F^\prime}_{21}(m_\pi^2,m_\pi^2,m_\pi^2;m_\pi^2)   ( 3 m_\pi^4 )
\nonumber\\&&
       + H^{F^\prime}_{21}(m_\pi^2,m_K^2,m_K^2;m_\pi^2)   (  - 3/8\, m_\pi^4 )
\nonumber\\&&
       + H^{F^\prime}_{21}(m_K^2,m_\pi^2,m_K^2;m_\pi^2)   ( 3 m_\pi^4 )
\nonumber\\&&
       + H^{F^\prime}_{21}(m_\eta^2,m_K^2,m_K^2;m_\pi^2)   ( 9/8\, m_\pi^4 )
\nonumber\\&&
       + H^F(m_\pi^2,m_\pi^2,m_\pi^2;m_\pi^2)   ( 5/3\, m_\pi^2 + 1/18\, q^2 )
\nonumber\\&&
       + H^F(m_\pi^2,m_K^2,m_K^2;m_\pi^2)   ( 1/12\, m_\pi^2 + 1/12\, q^2 )
\nonumber\\&&
       + H^F(m_K^2,m_\pi^2,m_K^2;m_\pi^2)   ( 15/32\, m_\pi^2 + 5/96\, (2 m_K^2 -m_\pi^2) - 5/48\, 
        q^2 )
\nonumber\\&&
       + H^F(m_K^2,m_K^2,m_\pi^2;m_\pi^2)   (  - 5/48\, m_K^2 )
\nonumber\\&&
       + H^F(m_K^2,m_K^2,m_\eta^2;m_\pi^2)   (  - 1/2\, m_\pi^2 - 1/16\, q^2 )
\nonumber\\&&
       + H^F_{1}(m_\pi^2,m_\pi^2,m_\pi^2;m_\pi^2)   (  - 3 m_\pi^2 - 1/3\, q^2 )
\nonumber\\&&
       + H^F_{1}(m_\pi^2,m_K^2,m_K^2;m_\pi^2)   ( 1/24\, q^2 )
\nonumber\\&&
       + H^F_{1}(m_K^2,m_\pi^2,m_K^2;m_\pi^2)   (  - m_\pi^2 - 1/8\, q^2 )
\nonumber\\&&
       + H^F_{1}(m_K^2,m_K^2,m_\eta^2;m_\pi^2)   ( m_\pi^2 + 1/8\, q^2 )
\nonumber\\&&
       + H^F_{21}(m_\pi^2,m_\pi^2,m_\pi^2;m_\pi^2)   ( 3 m_\pi^2 + 1/6\, q^2 )
\nonumber\\&&
       + H^F_{21}(m_\pi^2,m_K^2,m_K^2;m_\pi^2)   (  - 3/8\, m_\pi^2 - 1/48\, q^2 )
\nonumber\\&&
       + H^F_{21}(m_K^2,m_\pi^2,m_K^2;m_\pi^2)   ( 53/16\, m_\pi^2 + 1/16\, q^2 )
\nonumber\\&&
       + H^F_{21}(m_K^2,m_K^2,m_\pi^2;m_\pi^2)   (  - 5/16\, m_\pi^2 + 5/48\, q^2 )
\nonumber\\&&
       + H^F_{21}(m_\eta^2,m_K^2,m_K^2;m_\pi^2)   ( 9/8\, m_\pi^2 + 1/16\, q^2 )\,.
\end{eqnarray}
 
Secondly we have the contribution from diagram (b) in Fig.~\ref{fig3}. Most terms in the diagram can not be
reduced to the previous set of integrals and an independent basis is needed. In terms of these
we obtain the following result
\begin{eqnarray}
   F_{V\mathbf{V}}^\pi =&&
        V_{11}(m_\pi^2,m_\pi^2,m_\pi^2,m_\pi^2;m_\pi^2,q^2,m_\pi^2)   ( 5/2\, m_\pi^4 - 7/3\, 
         q^2 m_\pi^2 )
\nonumber\\&&
       + V_{11}(m_\pi^2,m_\pi^2,m_K^2,m_K^2;m_\pi^2,q^2,m_\pi^2)   ( m_\pi^4 - 2/3\, 
         q^2 m_\pi^2 + 1/12 q^4 )
\nonumber\\&&
       + V_{11}(m_\pi^2,m_\pi^2,m_\eta^2,m_\eta^2;m_\pi^2,q^2,m_\pi^2)   ( 1/18\, m_\pi^4 )
\nonumber\\&&
       + V_{11}(m_K^2,m_K^2,m_\pi^2,m_K^2;m_\pi^2,q^2,m_\pi^2)   ( 3/2\, m_\pi^4 - 17/12\, 
         q^2 m_\pi^2 + 1/6\, q^4 )
\nonumber\\&&
       + V_{11}(m_K^2,m_K^2,m_K^2,m_\eta^2;m_\pi^2,q^2,m_\pi^2)   ( 2/3\, m_\pi^4 - 2/3\, 
         q^2 m_\pi^2 + 1/8\, q^4 )
\nonumber\\&&
       + V_{21}(m_\pi^2,m_\pi^2,m_\pi^2,m_\pi^2;m_\pi^2,q^2,m_\pi^2)   (  - 6 m_\pi^2 + 
         q^2 )
\nonumber\\&&
       + V_{21}(m_\pi^2,m_\pi^2,m_K^2,m_K^2;m_\pi^2,q^2,m_\pi^2)   (  - 2 m_\pi^2 + 2/3\, 
         q^2 )
\nonumber\\&&
       + V_{21}(m_K^2,m_K^2,m_\pi^2,m_K^2;m_\pi^2,q^2,m_\pi^2)   (  - 4 m_\pi^2 + 4/3\, 
         q^2 )
\nonumber\\&&
       + V_{21}(m_K^2,m_K^2,m_K^2,m_\eta^2;m_\pi^2,q^2,m_\pi^2)   (  - 2 m_\pi^2 + 
         q^2 )
\nonumber\\&&
       + V_{22}(m_\pi^2,m_\pi^2,m_\pi^2,m_\pi^2;m_\pi^2,q^2,m_\pi^2)   (  - 6 m_\pi^4 + 10/3\, 
         q^2 m_\pi^2 )
\nonumber\\&&
       + V_{22}(m_\pi^2,m_\pi^2,m_K^2,m_K^2;m_\pi^2,q^2,m_\pi^2)   (  - 2 m_\pi^4 + 4/3\, 
         q^2 m_\pi^2 - 1/6\, q^4 )
\nonumber\\&&
       + V_{22}(m_K^2,m_K^2,m_\pi^2,m_K^2;m_\pi^2,q^2,m_\pi^2)   (  - 4 m_\pi^4 + 11/4\, 
         q^2 m_\pi^2 - 1/3\, q^4 )
\nonumber\\&&
       + V_{22}(m_K^2,m_K^2,m_K^2,m_\eta^2;m_\pi^2,q^2,m_\pi^2)   (  - 2 m_\pi^4 + 5/3\, 
         q^2 m_\pi^2 - 1/4\, q^4 )
\nonumber\\&&
       + V_{24}(m_\pi^2,m_\pi^2,m_\pi^2,m_\pi^2;m_\pi^2,q^2,m_\pi^2)   ( 5/3\, q^2 m_\pi^2
          + 1/2\, q^4 )
\nonumber\\&&
       + V_{24}(m_\pi^2,m_\pi^2,m_K^2,m_K^2;m_\pi^2,q^2,m_\pi^2)   ( 1/3\, q^2 m_\pi^2
          )
\nonumber\\&&
       + V_{24}(m_K^2,m_K^2,m_\pi^2,m_K^2;m_\pi^2,q^2,m_\pi^2)   ( 5/6\, q^2 m_\pi^2
          )
\nonumber\\&&
       + V_{24}(m_K^2,m_K^2,m_K^2,m_\eta^2;m_\pi^2,q^2,m_\pi^2)   ( 1/3\, q^2 m_\pi^2
          )
\nonumber\\&&
       + V_{25}(m_\pi^2,m_\pi^2,m_\pi^2,m_\pi^2;m_\pi^2,q^2,m_\pi^2)   (  - 4 m_\pi^2 + 4/3 \,
         q^2 )
\nonumber\\&&
       + V_{25}(m_\pi^2,m_\pi^2,m_K^2,m_K^2;m_\pi^2,q^2,m_\pi^2)   (  - 2 m_\pi^2 + 1/2 \,
         q^2 )
\nonumber\\&&
       + V_{25}(m_K^2,m_K^2,m_\pi^2,m_K^2;m_\pi^2,q^2,m_\pi^2)   (  - 3 m_\pi^2 + 17/12 \,
         q^2 )
\nonumber\\&&
       + V_{25}(m_K^2,m_K^2,m_K^2,m_\eta^2;m_\pi^2,q^2,m_\pi^2)   (  - 2 m_\pi^2 + 
         q^2 )
\nonumber\\&&
       + V_{26}(m_\pi^2,m_\pi^2,m_\pi^2,m_\pi^2;m_\pi^2,q^2,m_\pi^2)   (  - 4 m_\pi^4 + 7/3\, 
         q^2 m_\pi^2 + 1/3\, q^4 )
\nonumber\\&&
       + V_{26}(m_\pi^2,m_\pi^2,m_K^2,m_K^2;m_\pi^2,q^2,m_\pi^2)   (  - 2 m_\pi^4 + 
         q^2 m_\pi^2 )
\nonumber\\&&
       + V_{26}(m_K^2,m_K^2,m_\pi^2,m_K^2;m_\pi^2,q^2,m_\pi^2)   (  - 3 m_\pi^4 + 13/6\, 
         q^2 m_\pi^2 - 5/24\, q^4 )
\nonumber\\&&
       + V_{26}(m_K^2,m_K^2,m_K^2,m_\eta^2;m_\pi^2,q^2,m_\pi^2)   (  - 2 m_\pi^4 + 3/2\, 
         q^2 m_\pi^2 - 1/8\, q^4 )
\nonumber\\&&
       + V_{29}(m_\pi^2,m_\pi^2,m_\pi^2,m_\pi^2;m_\pi^2,q^2,m_\pi^2)   ( 4/3\, q^4 )
\nonumber\\&&
       + V_{29}(m_\pi^2,m_\pi^2,m_K^2,m_K^2;m_\pi^2,q^2,m_\pi^2)   ( 1/4\, q^4 )
\nonumber\\&&
       + V_{29}(m_K^2,m_K^2,m_\pi^2,m_K^2;m_\pi^2,q^2,m_\pi^2)   ( 7/24\, q^4 )
\nonumber\\&&
       + V_{29}(m_K^2,m_K^2,m_K^2,m_\eta^2;m_\pi^2,q^2,m_\pi^2)   ( 1/4\, q^4 )
\nonumber\\&&
       + V_{31}(m_\pi^2,m_\pi^2,m_\pi^2,m_\pi^2;m_\pi^2,q^2,m_\pi^2)   ( 6 m_\pi^2 - 2 
         q^2 )
\nonumber\\&&
       + V_{31}(m_\pi^2,m_\pi^2,m_K^2,m_K^2;m_\pi^2,q^2,m_\pi^2)   ( 3 m_\pi^2 - q^2
          )
\nonumber\\&&
       + V_{31}(m_K^2,m_K^2,m_\pi^2,m_K^2;m_\pi^2,q^2,m_\pi^2)   ( 6 m_\pi^2 - 2 
         q^2 )
\nonumber\\&&
       + V_{31}(m_K^2,m_K^2,m_K^2,m_\eta^2;m_\pi^2,q^2,m_\pi^2)   ( 9/2\, m_\pi^2 - 3/2\, 
         q^2 )
\nonumber\\&&
       + V_{32}(m_\pi^2,m_\pi^2,m_K^2,m_K^2;m_\pi^2,q^2,m_\pi^2)   (  - 1/3\, q^2 )
\nonumber\\&&
       + V_{32}(m_K^2,m_K^2,m_\pi^2,m_K^2;m_\pi^2,q^2,m_\pi^2)   (  - 2/3\, q^2 )
\nonumber\\&&
       + V_{32}(m_K^2,m_K^2,m_K^2,m_\eta^2;m_\pi^2,q^2,m_\pi^2)   (  - 1/2\, q^2 )
\nonumber\\&&
       + V_{33}(m_\pi^2,m_\pi^2,m_\pi^2,m_\pi^2;m_\pi^2,q^2,m_\pi^2)   ( 2 m_\pi^4 - 
         q^2 m_\pi^2 )
\nonumber\\&&
       + V_{33}(m_\pi^2,m_\pi^2,m_K^2,m_K^2;m_\pi^2,q^2,m_\pi^2)   ( m_\pi^4 - 2/3\, 
         q^2 m_\pi^2 + 1/12\, q^4 )
\nonumber\\&&
       + V_{33}(m_K^2,m_K^2,m_\pi^2,m_K^2;m_\pi^2,q^2,m_\pi^2)   ( 2 m_\pi^4 - 4/3\, 
         q^2 m_\pi^2 + 1/6\, q^4 )
\nonumber\\&&
       + V_{33}(m_K^2,m_K^2,m_K^2,m_\eta^2;m_\pi^2,q^2,m_\pi^2)   ( 3/2\, m_\pi^4 - 
         q^2 m_\pi^2 + 1/8\, q^4 )
\nonumber\\&&
       + V_{35}(m_\pi^2,m_\pi^2,m_\pi^2,m_\pi^2;m_\pi^2,q^2,m_\pi^2)   (  - 1/2\, q^4
          )
\nonumber\\&&
       + V_{35}(m_\pi^2,m_\pi^2,m_K^2,m_K^2;m_\pi^2,q^2,m_\pi^2)   (  - 1/3\, q^2 
         m_\pi^2 )
\nonumber\\&&
       + V_{35}(m_K^2,m_K^2,m_\pi^2,m_K^2;m_\pi^2,q^2,m_\pi^2)   (  - 2/3\, q^2 
         m_\pi^2 )
\nonumber\\&&
       + V_{35}(m_K^2,m_K^2,m_K^2,m_\eta^2;m_\pi^2,q^2,m_\pi^2)   (  - 1/2\, q^2 
         m_\pi^2 )
\nonumber\\&&
       + V_{36}(m_\pi^2,m_\pi^2,m_\pi^2,m_\pi^2;m_\pi^2,q^2,m_\pi^2)   (  - 1/2\, q^4
          )
\nonumber\\&&
       + V_{36}(m_\pi^2,m_\pi^2,m_K^2,m_K^2;m_\pi^2,q^2,m_\pi^2)   (  - 1/12\, q^4
          )
\nonumber\\&&
       + V_{36}(m_K^2,m_K^2,m_\pi^2,m_K^2;m_\pi^2,q^2,m_\pi^2)   (  - 1/6\, q^4
          )
\nonumber\\&&
       + V_{36}(m_K^2,m_K^2,m_K^2,m_\eta^2;m_\pi^2,q^2,m_\pi^2)   (  - 1/8\, q^4
          )
\nonumber\\&&
       + V_{37}(m_\pi^2,m_\pi^2,m_\pi^2,m_\pi^2;m_\pi^2,q^2,m_\pi^2)   ( 4 m_\pi^2 )
\nonumber\\&&
       + V_{37}(m_\pi^2,m_\pi^2,m_K^2,m_K^2;m_\pi^2,q^2,m_\pi^2)   ( 2 m_\pi^2 - 1/2\, 
         q^2 )
\nonumber\\&&
       + V_{37}(m_K^2,m_K^2,m_\pi^2,m_K^2;m_\pi^2,q^2,m_\pi^2)   ( 4 m_\pi^2 - q^2
          )
\nonumber\\&&
       + V_{37}(m_K^2,m_K^2,m_K^2,m_\eta^2;m_\pi^2,q^2,m_\pi^2)   ( 3 m_\pi^2 - 3/4\, 
         q^2 )
\nonumber\\&&
       + V_{38}(m_\pi^2,m_\pi^2,m_\pi^2,m_\pi^2;m_\pi^2,q^2,m_\pi^2)   ( 2 q^2 )
\nonumber\\&&
       + V_{39}(m_\pi^2,m_\pi^2,m_\pi^2,m_\pi^2;m_\pi^2,q^2,m_\pi^2)   ( 8 m_\pi^2 - 4 
         q^2 )
\nonumber\\&&
       + V_{39}(m_\pi^2,m_\pi^2,m_K^2,m_K^2;m_\pi^2,q^2,m_\pi^2)   ( 4 m_\pi^2 - 3/2\, 
         q^2 )
\nonumber\\&&
       + V_{39}(m_K^2,m_K^2,m_\pi^2,m_K^2;m_\pi^2,q^2,m_\pi^2)   ( 8 m_\pi^2 - 3 
         q^2 )
\nonumber\\&&
       + V_{39}(m_K^2,m_K^2,m_K^2,m_\eta^2;m_\pi^2,q^2,m_\pi^2)   ( 6 m_\pi^2 - 9/4\, 
         q^2 )
\nonumber\\&&
       + V_{310}(m_\pi^2,m_\pi^2,m_\pi^2,m_\pi^2;m_\pi^2,q^2,m_\pi^2)   (  - 2/3\, q^2 )
\nonumber\\&&
       + V_{310}(m_K^2,m_K^2,m_\pi^2,m_K^2;m_\pi^2,q^2,m_\pi^2)   (  - 5/6\, q^2 )
\nonumber\\&&
       + V_{310}(m_K^2,m_K^2,m_K^2,m_\eta^2;m_\pi^2,q^2,m_\pi^2)   (  - 1/2\, q^2 )
\nonumber\\&&
       + V_{311}(m_\pi^2,m_\pi^2,m_\pi^2,m_\pi^2;m_\pi^2,q^2,m_\pi^2)   ( 4 m_\pi^4 - 4/3\, 
         q^2 m_\pi^2 - 1/3\, q^4 )
\nonumber\\&&
       + V_{311}(m_\pi^2,m_\pi^2,m_K^2,m_K^2;m_\pi^2,q^2,m_\pi^2)   ( 2 m_\pi^4 - 
         q^2 m_\pi^2 )
\nonumber\\&&
       + V_{311}(m_K^2,m_K^2,m_\pi^2,m_K^2;m_\pi^2,q^2,m_\pi^2)   ( 4 m_\pi^4 - 29/12\, 
         q^2 m_\pi^2 + 5/24\, q^4 )
\nonumber\\&&
       + V_{311}(m_K^2,m_K^2,m_K^2,m_\eta^2;m_\pi^2,q^2,m_\pi^2)   ( 3 m_\pi^4 - 7/4\, 
         q^2 m_\pi^2 + 1/8\, q^4 )
\nonumber\\&&
       + V_{313}(m_\pi^2,m_\pi^2,m_\pi^2,m_\pi^2;m_\pi^2,q^2,m_\pi^2)   ( 2 q^2 m_\pi^2
          - 4/3\, q^4 )
\nonumber\\&&
       + V_{313}(m_\pi^2,m_\pi^2,m_K^2,m_K^2;m_\pi^2,q^2,m_\pi^2)   (  - 1/4\, q^4
          )
\nonumber\\&&
       + V_{313}(m_K^2,m_K^2,m_\pi^2,m_K^2;m_\pi^2,q^2,m_\pi^2)   (  - 7/24\, 
         q^4 )
\nonumber\\&&
       + V_{313}(m_K^2,m_K^2,m_K^2,m_\eta^2;m_\pi^2,q^2,m_\pi^2)   (  - 1/4\, q^4
          )
\nonumber\\&&
       + V_{315}(m_\pi^2,m_\pi^2,m_\pi^2,m_\pi^2;m_\pi^2,q^2,m_\pi^2)   (  - 2/3\, q^2 
         m_\pi^2 - 2/3\, q^4 )
\nonumber\\&&
       + V_{315}(m_\pi^2,m_\pi^2,m_K^2,m_K^2;m_\pi^2,q^2,m_\pi^2)   (  - 1/4\, q^4
          )
\nonumber\\&&
       + V_{315}(m_K^2,m_K^2,m_\pi^2,m_K^2;m_\pi^2,q^2,m_\pi^2)   (  - 5/6\, q^2 
         m_\pi^2 - 1/12\, q^4 )
\nonumber\\&&
       + V_{315}(m_K^2,m_K^2,m_K^2,m_\eta^2;m_\pi^2,q^2,m_\pi^2)   (  - 1/2\, q^2 
         m_\pi^2 - 1/8\, q^4 )
\nonumber\\&&
       + V_{316}(m_\pi^2,m_\pi^2,m_\pi^2,m_\pi^2;m_\pi^2,q^2,m_\pi^2)   (  - 5/3\, q^4
          )
\nonumber\\&&
       + V_{316}(m_\pi^2,m_\pi^2,m_K^2,m_K^2;m_\pi^2,q^2,m_\pi^2)   (  - 1/2\, q^4
          )
\nonumber\\&&
       + V_{316}(m_K^2,m_K^2,m_\pi^2,m_K^2;m_\pi^2,q^2,m_\pi^2)   (  - 7/12\, 
         q^4 )
\nonumber\\&&
       + V_{316}(m_K^2,m_K^2,m_K^2,m_\eta^2;m_\pi^2,q^2,m_\pi^2)   (  - 1/2\, q^4
          )
\nonumber\\&&
       + V_{317}(m_\pi^2,m_\pi^2,m_\pi^2,m_\pi^2;m_\pi^2,q^2,m_\pi^2)   ( 4 m_\pi^2 - 2 
         q^2 )
\nonumber\\&&
       + V_{317}(m_\pi^2,m_\pi^2,m_K^2,m_K^2;m_\pi^2,q^2,m_\pi^2)   ( 2 m_\pi^2 - q^2
          )
\nonumber\\&&
       + V_{317}(m_K^2,m_K^2,m_\pi^2,m_K^2;m_\pi^2,q^2,m_\pi^2)   ( 3/2\, m_\pi^2 - 3/4\, 
         q^2 )
\nonumber\\&&
       + V_{317}(m_K^2,m_K^2,m_K^2,m_\eta^2;m_\pi^2,q^2,m_\pi^2)   ( 3/2\, m_\pi^2 - 3/4\, 
         q^2 )
\nonumber\\&&
       + V_{319}(m_\pi^2,m_\pi^2,m_\pi^2,m_\pi^2;m_\pi^2,q^2,m_\pi^2)   ( 8 m_\pi^2 - 2 
         q^2 )
\nonumber\\&&
       + V_{(319}(m_\pi^2,m_\pi^2,m_K^2,m_K^2;m_\pi^2,q^2,m_\pi^2)   ( 4 m_\pi^2 - q^2
          )
\nonumber\\&&
       + V_{319}(m_K^2,m_K^2,m_\pi^2,m_K^2;m_\pi^2,q^2,m_\pi^2)   ( 3 m_\pi^2 - 3/4\, 
         q^2 )
\nonumber\\&&
       + V_{319}(m_K^2,m_K^2,m_K^2,m_\eta^2;m_\pi^2,q^2,m_\pi^2)   ( 3 m_\pi^2 - 3/4\, 
         q^2 )
\nonumber\\&&
       + V_{321}(m_\pi^2,m_\pi^2,m_\pi^2,m_\pi^2;m_\pi^2,q^2,m_\pi^2)   ( 4 m_\pi^4 - 2 
         q^2 m_\pi^2 )
\nonumber\\&&
       + V_{321}(m_\pi^2,m_\pi^2,m_K^2,m_K^2;m_\pi^2,q^2,m_\pi^2)   ( 2 m_\pi^4 - 
         q^2 m_\pi^2 )
\nonumber\\&&
       + V_{321}(m_K^2,m_K^2,m_\pi^2,m_K^2;m_\pi^2,q^2,m_\pi^2)   ( 3/2\, m_\pi^4 - 3/4\, 
         q^2 m_\pi^2 )
\nonumber\\&&
       + V_{321}(m_K^2,m_K^2,m_K^2,m_\eta^2;m_\pi^2,q^2,m_\pi^2)   ( 3/2\, m_\pi^4 - 3/4\, 
         q^2 m_\pi^2 )
\nonumber\\&&
       + V_{323}(m_\pi^2,m_\pi^2,m_\pi^2,m_\pi^2;m_\pi^2,q^2,m_\pi^2)   (  - q^4 )
\nonumber\\&&
       + V_{323}(m_\pi^2,m_\pi^2,m_K^2,m_K^2;m_\pi^2,q^2,m_\pi^2)   (  - 1/2\, q^4
          )
\nonumber\\&&
       + V_{323}(m_K^2,m_K^2,m_\pi^2,m_K^2;m_\pi^2,q^2,m_\pi^2)   (  - 3/8\, q^4
          )
\nonumber\\&&
       + V_{323}(m_K^2,m_K^2,m_K^2,m_\eta^2;m_\pi^2,q^2,m_\pi^2)   (  - 3/8\, q^4
          )
\nonumber\\&&
       + V_{325}(m_\pi^2,m_\pi^2,m_\pi^2,m_\pi^2;m_\pi^2,q^2,m_\pi^2)   (  - q^4 )
\nonumber\\&&
       + V_{325}(m_\pi^2,m_\pi^2,m_K^2,m_K^2;m_\pi^2,q^2,m_\pi^2)   (  - 1/2\, q^4
          )
\nonumber\\&&
       + V_{325}(m_K^2,m_K^2,m_\pi^2,m_K^2;m_\pi^2,q^2,m_\pi^2)   (  - 3/8\, q^4
          )
\nonumber\\&&
       + V_{325}(m_K^2,m_K^2,m_K^2,m_\eta^2;m_\pi^2,q^2,m_\pi^2)   (  - 3/8\, q^4
          )\,,
\end{eqnarray}
where the notation can be read from App.~\ref{int}.

\section{One-loop integrals}
\setcounter{equation}{0}
\label{appintegrals}

Through the calculation one has to use one-loop integrals of one, two and three
point functions. The latter disappear
after mass renormalization and the use of some
recursion relations. We only have to deal with
the following set of functions
-- in the remainder we use $d=4-2 \epsilon$.
\ba
\label{oneloop}
A(m_1^2)& =& \frac{1}{i}\int \frac{d^dq}{(2\pi)^d} \frac{1}{q^2-m_1^2}\,,\nonumber\\
B(m_1^2,m_2^2,p^2) &=& \frac{1}{i}\int 
\frac{d^dq}{(2\pi)^d} \frac{1}{(q^2-m_1^2)((q-p)^2-m_2^2)}\,,\nonumber\\
B_\mu(m_1^2,m_2^2,p^2) &=& \frac{1}{i}\int 
\frac{d^dq}{(2\pi)^d} \frac{q_\mu}{(q^2-m_1^2)((q-p)^2-m_2^2)}\nonumber\\
&=& p_\mu B_1(m_1^2,m_2^2,p^2)\,,\nonumber\\
B_{\mu\nu}(m_1^2,m_2^2,p^2) &=& \frac{1}{i}\int 
\frac{d^dq}{(2\pi)^d} \frac{q_\mu q_\nu}{(q^2-m_1^2)((q-p)^2-m_2^2)}\nonumber\\
&=& p_\mu p_\nu B_{21}(m_1^2,m_2^2,p^2)+g_{\mu \nu} B_{22}(m_1^2,m_2^2,p^2)\,,\nonumber\\
B_{\mu\nu\alpha}(m_1^2,m_2^2,p^2) &=& \frac{1}{i}\int 
\frac{d^dq}{(2\pi)^d} \frac{q_\mu q_\nu q_\alpha}{(q^2-m_1^2)((q-p)^2-m_2^2)}\nonumber\\
&& \hspace{-2cm}
= p_\mu p_\nu p_\alpha B_{31}(m_1^2,m_2^2,p^2) + 
(p_\mu g_{\nu \alpha} + p_\nu g_{\mu \alpha} + p_\alpha g_{\mu \nu})B_{32}(m_1^2,m_2^2,p^2)
\,.\nonumber\\
\ea
An expansion in $\epsilon$ leads to the following series
\ba
\label{expan}
A(m_1^2)& =& \frac{m_1^2}{16\pi^2}\lambda_0+\overline{A}(m_1^2)+\epsilon 
\overline{A}^\epsilon(m_1^2)+\ldots\,,
\nonumber\\
B_{ij}(m_1^2,m_2^2,s_\pi)& =& \frac{1}{16\pi^2} \mbox{pole}_{ij} 
+ \overline{B}_{ij}(m_1^2,m_2^2,s_\pi)
+\epsilon \overline{B}_{ij}^\epsilon(m_1^2,m_2^2,s_\pi) +\ldots\,,
\ea
with $\overline{A}, \overline{B}_{ij}$ defining finite quantities and
where "$\mbox{pole}_{ij}$" denotes the singular part of each of the $B_{ij}$ functions,
\ba
&&\mbox{pole} = \lambda_0\,,\quad
\mbox{pole}_{1} = \frac{\lambda_0}{2}\,,\quad 
\mbox{pole}_{21} = \frac{\lambda_0}{3}\,,\quad
\mbox{pole}_{22} = \frac{\lambda_0}{4}(m_1^2+m_2^2-\frac{s_\pi}{3})\,,\quad 
\nonumber\\&&
\mbox{pole}_{31} = \frac{\lambda_0}{4}\,,\quad
\mbox{pole}_{32} = \frac{\lambda_0}{24}(2m_1+4m_2^2-s_\pi)\,,
\ea
with
\be
\label{pole}
\lambda_0 = \frac{1}{\epsilon}+\ln(4\pi)+1-\gamma\,.
\ee

After some simpler algebraic manipulation, the functions defined in Eq. (\ref{oneloop})
can be related to the basic integrals $A(m_1^2)$ and $B_1(m_1^2,m_2^2,p^2)$ through the identities

\ba
\label{relations}
B_{31}(m_1^2,m_2^2,s_\pi)& =& \frac{1}{2s_\pi} \Big(
A(m_2^2)
     -(m_2^2-m_1^2-s_\pi)B_{21}(m_1^2,m_2^2,s_\pi)
-4B_{32}(m_1^2,m_2^2,s_\pi) \Big) \,, \nonumber\\ 
B_{32}(m_1^2,m_2^2,s_\pi)& =& \frac{1}{2s_\pi}
\Big( -\frac{m_1^2}{d}A(m_1^2)+\frac{m_2^2}{d}A(m_2^2)
-(m_2^2-m_1^2-s_\pi)B_{22}(m_1^2,m_2^2,s_\pi) \Big) \,,\nonumber\\
B_{21}(m_1^2,m_2^2,s_\pi)& =& 
\frac{1}{s_\pi}\Big( A(m_2^2) 
+m_1^2 B(m_1^2,m_2^2,s_\pi)
-d B_{22}(m_1^2,m_2^2,s_\pi)\Big) \,, \nonumber\\
B_{22}(m_1^2,m_2^2,s_\pi)& =& \frac{1}{2(d-1)} \Big(
A(m_2^2)+2m_1^2B(m_1^2,m_2^2,s_\pi)
\nonumber\\ &&
-(s_\pi+m_1^2-m_2^2)B_1(m_1^2,m_2^2,s_\pi) \Big)\,,  \nonumber\\
B_{1}(m_1^2,m_2^2,s_\pi)& =& \frac{1}{2s_\pi} \Big(
A(m_2^2)-A(m_1^2) 
   + ( m_1^2-m_2^2+ s_\pi ) B(m_1^2,m_2^2,s_\pi) \Big)\,, \nonumber\\
B(m_1^2,m_1^2,0)& = & \frac{(d-2)}{2m_1^2}A(m_1^2)\,. 
\ea

Notice that the previous identities
are only used at the final numerical level in order to avoid
cancellations between different terms occurring in the form factors.
The explicit expressions are
\ba
\overline{A}(m_1^2) &=& - \frac{m_1^2}{16\pi^2} \ln(m_1^2)\,,\nonumber\\
\overline{B}(m_1^2,m_2^2,p^2) &=&
-\frac{1}{16\pi^2}
\frac{m_1^2\ln(m_1^2)-m_2^2\ln(m_2^2)}{m_1^2-m_2^2}\nonumber\\&&
\hspace{-0.5cm}
+\frac{1}{(32\pi^2)} \left(
2+\left(-\frac{\Delta}{p^2}+\frac{\Sigma}{\Delta}\right)\ln\frac{m_1^2}{m_2^2}
-\frac{\nu}{p^2}\ln\frac{(p^2+\nu)^2-\Delta^2}{(p^2-\nu)^2-\Delta^2} \right)\,,
\ea
with $\Delta=m_1^2-m_2^2$, $\Sigma=m_1^2+m_2^2$
and $\nu^2 = [p^2-(m_1+m_2)^2][p^2-(m_1-m_2)^2]$.
Similarly combining Eqs. (\ref{expan}) and (\ref{relations}) one can obtain 
the expressions of the  $\epsilon$ terms in the expansion as functions of
$\overline{A}^\epsilon$ and $\overline{B}^\epsilon$.
A straight forward calculation leads to
\ba
16\pi^2 \overline{A}^\epsilon(m_1^2)&=& m_1^2[\frac{C^2}{2}+\frac{1}{2}+\frac{\pi^2}{12}+\frac{1}{2}
\ln^2(m_1^2)-C\ln(m_1^2)]\,, \nonumber\\
16\pi^2 \overline{B}^\epsilon(m_1^2,m_2^2,s_\pi) &=& \frac{C^2}{2}-\frac{1}{2}+\frac{\pi^2}{12}
+(C-1) \overline{B}(m_1^2,m_2^2,s_\pi) +\frac{1}{2} \int_0^1 dx\, \ln^2(m^2)\,,\nonumber\\
\ea
with $C=\ln(4\pi)+1-\gamma$ and $m^2= x m_1^2+(1-x)m_2^2-x(1-x)s_\pi$.
The definition (\ref{oneloop}) is $\mu$ independent. The subtraction
procedure will always provide the scale $\mu$ in the correct fashion to
give all logarithms dimensionless arguments of the type $\ln(m^2/\mu^2)$.

\section{Vertex integrals}
\setcounter{equation}{0}
\label{int}

In this appendix we display the vertex type integrals. 
They were needed to calculate the 
$K_{\ell 4}$ form factors \cite{ABT2,ABT3}, 
but we never showed their definition nor
how they appear in the expressions. The pion vector form factor
has a manageable expression, that allow us to show them. It thus makes sense
to display their definition and evaluation here.

We define 
\begin{equation}
\label{defvertex}
\langle\langle X \rangle\rangle =
\frac{1}{i^2}\int\frac{d^dr}{(2\pi)^d}\frac{d^ds}{(2\pi)^d}
\frac{X}{(r^2-m_1^2)((r-q)^2-m_2^2)(s^2-m_3^2)((r+s-p)^2-m_4^2)}\;,
\end{equation}

In the calculation one encounters 
integrals of this type with up to three integrated momenta in the numerator.
We Lorentz decompose them into scalar functions.
All these functions, referred to as vertex integrals, depend on
$m_1^2,m_2^2,m_3^2,m_4^2,p^2,q^2,(p-q)^2,\mu^2$. The scale
$\mu$ appears always in connection with the $\mu$-independent integrals
as defined in (\ref{defvertex}) to make all arguments of logarithms
dimensionless.
In the following Lorentz decomposition
the dependence on these arguments inside the $V_i$ functions
has been suppressed for brevity.
\begin{eqnarray}
\label{lorentz}
V &=&\langle\langle 1 \rangle\rangle\,,  
\nonumber\\
 V_\mu^r &=&\langle\langle r_\mu \rangle\rangle 
= p_\mu V_{11} + q_\mu V_{12}\,,
\nonumber\\
 V_\mu^s&=&\langle\langle s_\mu \rangle\rangle 
=  p_\mu V_{13} + q_\mu V_{14}\,,
\nonumber\\
V_{\mu\nu}^{rr} &=& \langle\langle r_\mu r_\nu \rangle\rangle 
 = g_{\mu\nu}V_{21}
+p_\mu p_\nu V_{22} + q_\mu q_\nu V_{23}
+(p_\mu q_\nu + q_\mu p_\nu)V_{24}\,,
\nonumber\\
 V_{\mu\nu}^{rs} &=& \langle\langle r_\mu s_\nu \rangle\rangle 
 = g_{\mu\nu}V_{25}
+p_\mu p_\nu V_{26} + q_\mu q_\nu V_{27}
+q_\mu p_\nu V_{28}+ p_\mu q_\nu V_{29}\,,
\nonumber\\
V_{\mu\nu}^{ss} &=& \langle\langle r_\mu s_\nu \rangle\rangle 
 = g_{\mu\nu}V_{210}
+p_\mu p_\nu V_{211} + q_\mu q_\nu V_{212}
+(q_\mu p_\nu + p_\mu q_\nu) V_{213}\,,
\nonumber\\
V_{\mu\nu\alpha}^{rrr} &=& \langle\langle r_\mu r_\nu r_\alpha\rangle\rangle 
 = (g_{\mu\nu}p_\alpha+g_{\mu\alpha}p_\nu+g_{\nu\alpha}p_\mu)V_{31}
+(g_{\mu\nu}q_\alpha+g_{\mu\alpha}q_\nu+g_{\nu\alpha}q_\mu)V_{32}
\nonumber\\&&
+p_\mu p_\nu p_\alpha V_{33} + q_\mu q_\nu q_\alpha V_{34}
+(p_\mu p_\nu q_\alpha+ p_\mu q_\nu p_\alpha + q_\mu p_\nu p_\alpha)V_{35}
\nonumber\\&&
+(q_\mu q_\nu p_\alpha+ q_\mu p_\nu q_\alpha + p_\mu q_\nu q_\alpha)V_{36}\,,
\nonumber\\
V_{\mu\nu\alpha}^{rrs} &=& \langle\langle r_\mu r_\nu s_\alpha\rangle\rangle 
 = g_{\mu\nu}p_\alpha V_{37}+g_{\mu\nu}q_\alpha V_{38}
+(g_{\mu\alpha}p_\nu+g_{\nu\alpha}p_\mu)V_{39}
\nonumber\\&&
+(g_{\mu\alpha}q_\nu+g_{\nu\alpha}q_\mu)V_{310}
+p_\mu p_\nu p_\alpha V_{311} + q_\mu q_\nu q_\alpha V_{312}
+p_\mu p_\nu q_\alpha V_{313}
\nonumber\\&&
+ q_\mu q_\nu p_\alpha V_{314}
+(p_\mu q_\nu+q_\mu p_\nu)p_\alpha V_{315}
+(p_\mu q_\nu+q_\mu p_\nu)q_\alpha V_{316}\,,
\nonumber\\
V_{\mu\nu\alpha}^{rss} &=& \langle\langle r_\mu s_\nu s_\alpha\rangle\rangle 
 = p_\mu g_{\nu\alpha} V_{317}+q_\mu g_{\nu\alpha} V_{318}
+(g_{\mu\nu}p_\alpha+g_{\mu\alpha}p_\nu)V_{319}
\nonumber\\&&
+(g_{\mu\nu}q_\alpha+g_{\mu\alpha}q_\nu)V_{320}
+p_\mu p_\nu p_\alpha V_{321} + q_\mu q_\nu q_\alpha V_{322}
+p_\mu q_\nu q_\alpha V_{323}
\nonumber\\&&
+ q_\mu p_\nu p_\alpha V_{324}
+p_\mu(p_\nu q_\alpha+q_\nu p_\alpha) V_{325}
+q_\mu(p_\nu q_\alpha+q_\nu p_\alpha) V_{326}\,.
\nonumber\\
%
\end{eqnarray}

This set of 44 functions is not obviously a \emph{minimum}
set and 
some relations can be found between then. Instead of reducing to a basic set, just by simply
manipulations like contracting with the external momenta or a $g_{\mu\nu}$ tensor, we use
those as cross-check of our numerical integrals.
In addition one can relate functions with different arguments
with some redefinitions of momenta. For instance  
the substitution $r\to-r'+q;s\to-s'$ 
relates functions with the canonical arguments
$m_1^2,m_2^2,m_3^2,m_4^2,p^2,q^2,(p-q)^2,\mu^2$
to those with arguments
$m_2^2,m_1^2,m_3^2,m_4^2,(q-p)^2,q^2,p^2,\mu^2$, while
$r\to r';s\to-r'-s'+p$
relates them to those with arguments 
$m_1^2,m_2^2,m_4^2,m_3^2,p^2,q^2,(p-q)^2,\mu^2$. As an especial case we find a 
huge simplification in our results when we deal with the substitution
$r\to r';s\to-r'-s'+p$ for the more restrictive case $m_3^2=m_4^2$ obtaining the following
set of relations between the various integrals in Eq. (\ref{lorentz})
\ba
&& \hspace{-1.0cm}V_\mu^s = \frac{1}{2} (p_\mu V -V_\mu^r) \,, \quad
V_{\mu\nu}^{rs}= \frac{1}{2} (p_\nu V_\mu^r -V_{\mu\nu}^{rr}) \,, \quad
V_{\mu\nu\alpha}^{rrs}= \frac{1}{2} (p_\alpha V_{\mu\nu}^{rr} -V_{\mu\nu\alpha}^{rrr}) \,. 
\ea

In order to evaluate these functions we used the methods developed
in \cite{GB,GY}, where we refer the reader for a more extensive treatment. 
Here we only repeat the basic steps -- our functions differ
slightly from the ones in \cite{GB,GY} since we stay in Minkowski
space throughout.\\
We first combine the first two propagators with a Feynman integration in Eq.~(\ref{defvertex})

\begin{eqnarray}
\langle \langle X \rangle\rangle &=&
\int_0^1 dy 
\frac{1}{i^2}\int\frac{d^dr}{(2\pi)^d}\frac{d^ds}{(2\pi)^d}
\frac{X}{[(r-yq)^2-((1-y)m_1^2+y m_2^2-y(1-y)q^2)]^2}\times
\nonumber\\&&
\frac{1}{
(s^2-m_3^2)
 ((r-yq+s-(p-yq))^2-m_4^2)}\;,
\end{eqnarray}
and we shift the $r$ integration to $r-yq$. All remaining integrals
can be written in terms of 
\begin{equation}
P_{211}^{ij}(m^2,m_3^2,m_4^2,k^2) =
\frac{1}{i^2}\int\frac{d^dr}{(2\pi)^d}\frac{d^ds}{(2\pi)^d}
\frac{(r\cdot k)^i (s\cdot k)^j}{(r^2-m^2)^2(s^2-m_3^2)((r+s+k)^2-m_4^2)}
\end{equation}
where in our case $ij$ is reduced to the values
$00,10,01,20,11,02,30,21$ and $12$.
We also have defined 
\be
\label{variables}
m^2 = (1-y)m_1^2+y m_2^2-y(1-y)q^2\,,\quad 
k = -p+yq\,.
\ee
The infinite parts of these functions can be obtained analytically and their
finite parts can be written in terms of $9$ functions, $h_i$, that have a fairly
simple integral representation -- see \cite{GY} for an explicit representation of these functions.
So in the end all the functions appearing in the vertex diagrams are obtained
as a double integral. The singularities in these integrals can be
avoided by deforming the integration paths 
, this also serves as a check on the numerical code.

Following the previous steps is straight forward to obtain the relations between the vertex, 
$V_i$, and the \emph{basic} integrals, $P^{ij}_{abc}$, 
in terms of the variables defined in Eq. (\ref{variables}) -- in what follows and for simplicity
$V_i$ should be read $V_i(m_1^2,m_2^2,m_3^2,m_4^2;p^2,q^2,(p-q)^2)$. 
\ba
V &=& \Big\{ P^{00}_{211} \Big\}_y [m^2,m^2_3,m^2_4;k^2]\,,
\nonumber\\
V_{11}& =& \Big\{ - \frac{P^{10}_{211}}{k^2} \Big\}_y [m^2,m^2_3,m^2_4;k^2]\,,
\nonumber\\
V_{12}& =& \Big\{ y\Big( P^{00}_{211}
    - \frac{P^{10}_{211}}{k^2} \Big) \Big\}_y [m^2,m^2_3,m^2_4;k^2]\,,
\nonumber\\
V_{13}& =& \Big\{ - \frac{P^{01}_{211}}{k^2} \Big\}_y [m^2,m^2_3,m^2_4;k^2]\,,
\nonumber\\
V_{14}& =& \Big\{ y \frac{P^{01}_{211}}{k^2} \Big\}_y [m^2,m^2_3,m^2_4;k^2]\,,
\nonumber\\
V_{21}& =& \frac{1}{(n-1)} \Big\{
- \frac{P^{20}_{211}}{k^2} + P^{00}_{111} +m^2 P^{00}_{211}  \Big\}_y [m^2,m^2_3,m^2_4;k^2]\,,
\nonumber\\
V_{22}& =& \Big\{
\frac{1}{k^2(n-1)} (  \frac{P^{20}_{211}}{k^2} - P^{00}_{111} - m^2P^{00}_{211} )
+        \frac{P^{20}_{211}}{k^4} \Big\}_y [m^2,m^2_3,m^2_4;k^2]\,,
\nonumber\\
V_{23}& =& \Big\{
\frac{y^2}{k^2(n-1)}  ( \frac{P^{20}_{211}}{k^2} - P^{00}_{111} - m^2P^{00}_{211} )
       + y^2  ( \frac{P^{20}_{211}}{k^4} + 2 \frac{P^{10}_{211}}{k^2} + P^{00}_{211} ) 
\Big\}_y [m^2,m^2_3,m^2_4;k^2]\,,
\nonumber\\
V_{24}& =& \Big\{
\frac{y}{k^2(n-1)} (  -\frac{P^{20}_{211}}{k^2} + P^{00}_{111} + m^2P^{00}_{211} )
       - \frac{y}{k^2} (   \frac{P^{20}_{211}}{k^2} + P^{10}_{211} ) \Big\}_y [m^2,m^2_3,m^2_4;k^2]\,,
\nonumber\\
V_{25}& =& \frac{1}{2(n-1)} \Big\{
         - 2P^{10}_{211} - P^{00}_{111} - 2 P^{01}_{211} - 
         (k^2+m^2+m_3^2-m_4^2)  P^{00}_{211} 
\nonumber\\&&
         + B(m^2,m^2,0) (A(m_3^2)- A(m_4^2))
         - 2 \frac{P^{11}_{211}}{k^2} ) \Big\}_y [m^2,m^2_3,m^2_4;k^2]\,,
\nonumber\\
V_{26}& =& \Big\{
       \frac{1}{2 k^2(n-1)} \Big( 2 P^{10}_{211} + P^{00}_{111} + 2 P^{01}_{211}
          +(k^2 +m^2 + m_3^2 - m_4^2) P^{00}_{211}
\nonumber\\&&
        - B(m^2,m^2,0) (A(m_3^2)- A(m_4^2)) 
         + 2 \frac{P^{11}_{211}}{k^2} \Big) 
       + \frac{P^{11}_{211}}{k^4} \Big\}_y [m^2,m^2_3,m^2_4;k^2]\,,
\nonumber\\
V_{27}& =& \Big\{
\frac{y^2}{2 k^2(n-1)} ( 2 P^{10}_{211} + P^{00}_{111} + 2 P^{01}_{211}
         +(k^2+m^2 + m_3^2 - m_4^2) P^{00}_{211}  
\nonumber\\&&
         - B(m^2,m^2,0) (A(m_3^2)- A(m_4^2)) 
         + 2 \frac{P^{11}_{211}}{k^2} )
+\frac{y^2}{ k^2}   ( P^{01}_{211} + \frac{P^{11}_{211}}{k^2} ) \Big\}_y [m^2,m^2_3,m^2_4;k^2]\,,
\nonumber\\
V_{28}& =& \Big\{
\frac{y}{2 k^2(n-1)} \Big(  - 2 P^{10}_{211} - P^{00}_{111} - 2 P^{01}_{211}
         -( k^2 + m^2 + m_3^2 - m_4^2) P^{00}_{211} 
\nonumber\\&&
         + B(m^2,m^2,0) (A(m_3^2)- A(m_4^2))
         - 2 \frac{P^{11}_{211}}{k^2}  \Big)
-\frac{y}{ k^2}   
        (  P^{01}_{211} + \frac{P^{11}_{211}}{k^2} ) \Big\}_y [m^2,m^2_3,m^2_4;k^2]\,,
\nonumber\\
V_{29}& =& \Big\{
\frac{y}{2 k^2(n-1)} \Big(  - 2 P^{10}_{211} - P^{00}_{111}  - 2 P^{01}_{211}
         -(k^2 + m^2 + m_3^2 - m_4^2) P^{00}_{211}
\nonumber\\&&
         + B(m^2,m^2,0) (A(m_3^2)- A(m_4^2))
         - 2 \frac{P^{11}_{211}}{k^2}  \Big)
-\frac{y}{ k^4} P^{11}_{211} \Big\}_y [m^2,m^2_3,m^2_4;k^2]\,,
\nonumber\\
V_{210}& =& \frac{1}{(n-1)} \Big\{
 m_3^2 P^{00}_{211} + A(m_4^2)B(m^2,m^2,0) - \frac{P^{02}_{211}}{k^2}
\Big\}_y [m^2,m^2_3,m^2_4;k^2]\,,
\nonumber\\
V_{211}& =&  \Big\{
\frac{1}{k^2 (n-1)} \Big(  - m_3^2 P^{00}_{211}  - A(m_4^2)B(m^2,m^2,0)
          + \frac{P^{02}_{211}}{k^2} \Big)
       +  \frac{P^{02}_{211}}{k^4}  \Big\}_y [m^2,m^2_3,m^2_4;k^2]\,,
\nonumber\\
V_{212}& =&  \Big\{
\frac{y^2}{k^2 (n-1)} \Big( - m_3^2 P^{00}_{211}  - A(m_4^2)B(m^2,m^2,0) 
          + \frac{P^{02}_{211}}{k^2} \Big)
          + \frac{y^2}{k^4} P^{02}_{211} \Big\}_y [m^2,m^2_3,m^2_4;k^2]\,,
\nonumber\\
V_{213}& =&  \Big\{
\frac{y}{k^2 (n-1)}  \Big( m_3^2 P^{00}_{211}  + A(m_4^2)B(m^2,m^2,0) 
          - \frac{P^{02}_{211}}{k^2} \Big)
- \frac{y}{k^4} P^{02}_{211}  \Big\}_y [m^2,m^2_3,m^2_4;k^2]\,,
\nonumber\\
V_{31}& =&  \frac{1}{(n-1)} \Big\{
\frac{1}{k^2} (  - m^2 P^{10}_{211} - P^{10}_{111}
          + \frac{P^{30}_{211}}{k^2} ) \Big\}_y [m^2,m^2_3,m^2_4;k^2]\,,
\nonumber\\
V_{32}& =&  \frac{1}{(n-1)} \Big\{
 y \Big( m^2 P^{00}_{211} + P^{00}_{111}  + \frac{1}{k^2} ( m^2 P^{10}_{211} 
         + P^{10}_{111}  - P^{20}_{211}  - \frac{ P^{30}_{211}}{k^2})\Big) 
\Big\}_y [m^2,m^2_3,m^2_4;k^2]\,,
\nonumber\\
V_{33}& =&   \Big\{
        - \frac{ P^{30}_{211}}{k^6} 
+ \frac{3}{k^4(n-1)} ( m^2 P^{10}_{211}  + P^{10}_{111}  - 
         \frac{ P^{30}_{211}}{k^2}  ) \Big\}_y [m^2,m^2_3,m^2_4;k^2]\,,
\nonumber\\
V_{34}& =&   \Big\{
       y^3 \Big( P^{00}_{211}  + \frac{3}{k^2} P^{10}_{211}  + \frac{3}{k^4} P^{20}_{211} 
         + \frac{ P^{30}_{211}}{k^6}  
+ \frac{3}{k^2(n-1)}  (  - m^2 P^{00}_{211}  - P^{00}_{111} 
\nonumber\\&&
          + \frac{1}{k^2} ( - m^2 P^{10}_{211}  - P^{10}_{111}  + P^{20}_{211} 
          + P^{30}_{211}  ) ) \Big) \Big\}_y [m^2,m^2_3,m^2_4;k^2]\,,
\nonumber\\
V_{35}& =&   \Big\{
       y \Big( \frac{1}{k^4} ( P^{20}_{211}  + \frac{ P^{30}_{211}}{k^2}  )
+ \frac{1}{k^2(n-1)}  (  - m^2 P^{00}_{211}  - P^{00}_{111}  
\nonumber\\&&
+ \frac{1}{k^2} ( - 3m^2 P^{10}_{211} 
          - 3P^{10}_{111} + P^{20}_{211}  + 3 \frac{ P^{30}_{211}}{k^2} ))   \Big)
\Big\}_y [m^2,m^2_3,m^2_4;k^2]\,,
\nonumber\\
V_{36}& =&   \Big\{
        \frac{y^2}{k^2} \Big(  - P^{10}_{211}  - \frac{2}{k^2} P^{20}_{211}-\frac{ P^{30}_{211}}{k^4}
       + \frac{1}{(n-1)} ( 2 m^2 P^{00}_{211}  + 2P^{00}_{111}  
\nonumber\\&&
+ \frac{1}{k^2} ( 3m^2
         P^{10}_{211}  + 3P^{10}_{111}  - 2P^{20}_{211}  - 3
         \frac{ P^{30}_{211}}{k^2}  ) ) \Big) \Big\}_y [m^2,m^2_3,m^2_4;k^2]\,,
\nonumber\\
V_{37}& =&   \Big\{
 \frac{1}{k^2(n-1)}  (  - m^2 P^{01}_{211} -  P^{01}_{111}  + \frac{ P^{21}_{211}}{k^2} )
\Big\}_y [m^2,m^2_3,m^2_4;k^2]\,,
\nonumber\\
V_{38}& =&   \Big\{
\frac{y}{k^2(n-1)}  ( m^2 P^{01}_{211}  + P^{01}_{111}  - \frac{ P^{21}_{211}}{k^2} )
\Big\}_y [m^2,m^2_3,m^2_4;k^2]\,,
\nonumber\\
V_{39}& =&    \Big\{
\frac{1}{2k^2 (n-1)}  \Big( (k^2+m^2 +m_3^2 -m_4^2 ) P^{10}_{211} 
         + P^{10}_{111}  + 2 P^{11}_{211}  + 
         2 P^{20}_{211}  
\nonumber\\&&
+ 2 \frac{ P^{21}_{211}}{k^2}  \Big) \Big\}_y [m^2,m^2_3,m^2_4;k^2]\,,
\nonumber\\
V_{310}& =&    \Big\{
\frac{y}{2 (n-1)}    \Big( B(m^2,m^2,0) ( A(m_3^2) - A(m_4^2) )
          - (k^2+m^2 +m_3^2 -m_4^2 ) P^{00}_{211} 
\nonumber\\&&
         - P^{00}_{111}  - 2  P^{01}_{211} + \frac{1}{k^2} (  - ( m^2+m_3^2 +3 k^2 -m_4^2)  
         P^{10}_{211} - 
         P^{10}_{111} - 4 P^{11}_{211}  - 2 P^{20}_{211} 
\nonumber\\&&
- 2 \frac{ P^{21}_{211}}{k^2}  ) \Big)
\Big\}_y [m^2,m^2_3,m^2_4;k^2]\,,
\nonumber\\
V_{311}& =&    \Big\{ \frac{1}{k^4} \Big(
        - \frac{ P^{21}_{211}}{k^2} 
+ \frac{1}{(n-1)}  ( m^2 P^{01}_{211} + P^{01}_{111} - (k^2 +m^2 +m_3^2 -m_4^2)  
         P^{10}_{211} - P^{10}_{111} 
\nonumber\\&&
         - 2P^{11}_{211}  - 2P^{20}_{211} - 3 \frac{ P^{21}_{211}}{k^2} ) \Big)
\Big\}_y [m^2,m^2_3,m^2_4;k^2]\,,
\nonumber\\
V_{312}& =&    \Big\{ \frac{y^3}{k^2} \Big( 
        P^{01}_{211} + 2\frac{ P^{11}_{211}}{k^2}  + \frac{ P^{21}_{211}}{k^4}  
       + \frac{1}{(n-1)} (  B(m^2,m^2,0) ( A(m_4^2) - A(m_3^2) ) 
\nonumber\\&&
          + (k^2 + m^2 +m_3^2 -m_4^2) P^{00}_{211} 
         + P^{00}_{111}  + 2 P^{01}_{211} 
          - \frac{ m^2}{k^2} P^{01}_{211}- \frac{ P^{01}_{111}}{k^2} 
\nonumber\\&&
+ \frac{1}{k^2}( (3k^2 + 
         m^2 + m_3^2 - m_4^2) P^{10}_{211}
          + P^{10}_{111} + 4P^{11}_{211}  + 2P^{20}_{211}  
\nonumber\\&&
+ 3
         \frac{ P^{21}_{211}}{k^2}    ))\Big) \Big\}_y [m^2,m^2_3,m^2_4;k^2]\,,
\nonumber\\
V_{313}& =&    \Big\{ \frac{y}{k^4} 
        \Big( \frac{ P^{21}_{211}}{k^2} 
       + \frac{1}{(n-1)} (  - m^2 P^{01}_{211} - P^{01}_{111} + ( k^2+m^2+m_3^2-m_4^2) P^{10}_{211} 
         + P^{10}_{111}  
\nonumber\\&&
+ 2P^{11}_{211} + 2P^{20}_{211}  + 3\frac{ P^{21}_{211}}{k^2} 
          ) \Big) \Big\}_y [m^2,m^2_3,m^2_4;k^2]\,,
\nonumber\\
V_{314}& =&    \Big\{ \frac{y^2}{k^2} 
       \Big(  - P^{01}_{211}  - 2\frac{ P^{11}_{211}}{k^2}  - \frac{ P^{21}_{211}}{k^4}
+ \frac{1}{(n-1)}  ( B(m^2,m^2,0) ( A(m_3^2) - A(m_4^2))
\nonumber\\&&
          - ( k^2+m^2+m_3^2-m_4^2) P^{00}_{211}
         - P^{00}_{111}  - 2P^{01}_{211}
+\frac{1}{k^2} (
           m^2 P^{01}_{211}  + P^{01}_{111}  
\nonumber\\&&
- (3k^2+m^2 +m_3^2-m_4^2) P^{10}_{211}  
          - P^{10}_{111}  - 4P^{11}_{211}  - 2P^{20}_{211}  - 3 \frac{ P^{21}_{211}}{k^2}
          )) \Big) \Big\}_y [m^2,m^2_3,m^2_4;k^2]\,,
\nonumber\\
V_{315}& =&    \Big\{ \frac{y}{k^2} 
       \Big( \frac{ P^{11}_{211}}{k^2}  + \frac{ P^{21}_{211}}{k^4}  
       + \frac{1}{(n-1)} (   \frac{1}{2} B(m^2,m^2,0)( A(m_4^2)-A(m_3^2)) 
\nonumber\\&&
          + \frac{1}{2} (k^2 +m^2+m_3^2-m_4^2) P^{00}_{211} 
         + \frac{1}{2} P^{00}_{111}  + P^{01}_{211} 
+\frac{1}{k^2} (
          - m^2 P^{01}_{211}  - P^{01}_{111}  
\nonumber\\&&
+ (2k^2+m^2+m_3^2-m_4^2)P^{10}_{211}  
        + P^{10}_{111}  + 3P^{11}_{211}  + 2P^{20}_{211} + 3\frac{ P^{21}_{211}}{k^2} ) ) \Big)
\Big\}_y [m^2,m^2_3,m^2_4;k^2]\,,
\nonumber\\
V_{316}& =&    \Big\{ \frac{y^2}{k^2} 
        \Big(  - \frac{ P^{11}_{211}}{k^2} - \frac{ P^{21}_{211}}{k^4}  
       + \frac{1}{(n-1)} ( \frac{1}{2} B(m^2,m^2,0)( A(m_3^2) - A(m_4^2))
\nonumber\\&&
         - \frac{1}{2} (k^2 +m^2+m_3^2-m_4^2) P^{00}_{211} 
         - \frac{1}{2} P^{00}_{111} 
         - P^{01}_{211}  
+\frac{1}{k^2} (
        m^2 P^{01}_{211} + P^{01}_{111} 
\nonumber\\&&
- (2k^2 
         +m^2+m_3^2-m_4^2) P^{10}_{211}  
         - P^{10}_{111}  - 3P^{11}_{211}  - 2P^{20}_{211} 
          - 3\frac{ P^{21}_{211}}{k^2}  ) ) \Big) \Big\}_y [m^2,m^2_3,m^2_4;k^2]\,,
\nonumber\\ 
V_{317}& =& \frac{1}{(n-1)}    \Big\{ \frac{1}{k^2} 
 (  - m_3^2 P^{10}_{211} + \frac{ P^{12}_{211}}{k^2}  )
\Big\}_y [m^2,m^2_3,m^2_4;k^2]\,,
\nonumber\\
V_{318}& =& \frac{1}{(n-1)}   \Big\{ y 
        \Big( A(m_4^2) B(m^2,m^2,0) + m_3^2 P^{00}_{211} 
+ \frac{1}{k^2} (  - P^{02}_{211} 
          + m_3^2 P^{10}_{211}  
\nonumber\\&&
- \frac{ P^{12}_{211}}{k^2} ) \Big) \Big\}_y [m^2,m^2_3,m^2_4;k^2]\,,
\nonumber\\
V_{319}& =& \frac{1}{2 (n-1)}   \Big\{  
  - A(m_4^2) B(m^2,m^2,0) + \frac{1}{k^2}\Big(  (k^2 +m^2 +m_3^2-m_4^2) P^{01}_{211}
\nonumber\\&&
          + P^{01}_{111} 
          + 2 P^{02}_{211}  + 2 P^{11}_{211}  + 2 \frac{ P^{12}_{211}}{k^2}  \Big)
\Big\}_y [m^2,m^2_3,m^2_4;k^2]\,,
\nonumber\\
V_{320}& =& \frac{1}{2 (n-1)}   \Big\{ y \Big( 
  A(m_4^2) B(m^2,m^2,0) + \frac{1}{k^2} ( - (k^2+m^2+m_3^2-m_4^2) P^{01}_{211}  
\nonumber\\&&
         - P^{01}_{111}  - 2P^{02}_{211} - 2P^{11}_{211} - 2\frac{ P^{12}_{211}}{k^2} )\Big)
\Big\}_y [m^2,m^2_3,m^2_4;k^2]\,,
\nonumber\\
V_{321}& =&    \Big\{ \frac{1}{k^2} \Big(
        - \frac{ P^{12}_{211}}{k^4} 
       + \frac{1}{(n-1)}  ( A(m_4^2) B(m^2,m^2,0)  
+\frac{1}{k^2} (
- (k^2 +m^2+m_3^2-m_4^2) P^{01}_{211}  
\nonumber\\&&
        - P^{01}_{111}  - 2
         P^{02}_{211} + m_3^2 P^{10}_{211} - 2P^{11}_{211} - 3\frac{ P^{12}_{211}}{k^2} ) ) \Big)
\Big\}_y [m^2,m^2_3,m^2_4;k^2]\,,
\nonumber\\
V_{322}& =&    \Big\{  
   \frac{y^3}{ k^2} \Big(      \frac{ P^{02}_{211}}{k^2}  + \frac{ P^{12}_{211}}{k^4} 
       + \frac{1}{(n-1)} (  - 2A(m_4^2) B(m^2,m^2,0) - m_3^2 P^{00}_{211} 
\nonumber\\&&
         + \frac{1}{k^2} ( (k^2 +m^2+m_3^2-m_4^2) P^{01}_{211} 
         + P^{01}_{111}  + 3P^{02}_{211} 
          - m_3^2 P^{10}_{211}  + 2P^{11}_{211} 
\nonumber\\&&
+ 3\frac{ P^{12}_{211}}{k^2} ) ) \Big)
\Big\}_y [m^2,m^2_3,m^2_4;k^2]\,,
\nonumber\\
V_{323}& =&    \Big\{  \frac{y^2}{k^2} \Big(
          - \frac{ P^{12}_{211}}{k^4}
+ \frac{1}{(n-1)}  ( A(m_4^2) B(m^2,m^2,0)  
+ \frac{1}{k^2} ( - (k^2 +m^2+m_3^2-m_4^2) P^{01}_{211}  
\nonumber\\&&
          - P^{01}_{111}  - 2 P^{02}_{211}  + m_3^2 P^{10}_{211}  - 2
         P^{11}_{211}  - 3\frac{ P^{12}_{211}}{k^2}  ) ) \Big)\Big\}_y [m^2,m^2_3,m^2_4;k^2]\,,
\nonumber\\
V_{324}& =&    \Big\{  \frac{y}{k^2} \Big(
       \frac{ P^{02}_{211}}{k^2}   + \frac{ P^{12}_{211}}{k^4} 
       + \frac{1}{(n-1)} (  - 2A(m_4^2) B(m^2,m^2,0) - m_3^2 P^{00}_{211} 
\nonumber\\&&
+ \frac{1}{k^2} (  (k^2 +m^2+m_3^2-m_4^2) P^{01}_{211}
         + P^{01}_{111}  + 3P^{02}_{211}  - m_3^2 P^{10}_{211} 
         + 2P^{11}_{211} 
\nonumber\\&&
+ 3 \frac{ P^{12}_{211}}{k^2}  ) )\Big)\Big\}_y [m^2,m^2_3,m^2_4;k^2]\,,
\nonumber\\
V_{325}& =&    \Big\{ \frac{y}{k^2} \Big( 
        \frac{ P^{12}_{211}}{k^4}  
       + \frac{1}{(n-1)}  (  - A(m_4^2) B(m^2,m^2,0) + \frac{1}{k^2} (
(k^2 +m^2+m_3^2-m_4^2) P^{01}_{211}  
\nonumber\\&&
         + P^{01}_{111}  
+ 2P^{02}_{211}  - m_3^2 P^{10}_{211}  + 2P^{11}_{211} 
         + 3\frac{ P^{12}_{211}}{k^2} ) ) \Big) \Big\}_y [m^2,m^2_3,m^2_4;k^2]\,,
\nonumber\\
V_{326}& =&    \Big\{ \frac{y^2}{k^2} \Big( 
         - \frac{ P^{02}_{211}}{k^2} - \frac{ P^{12}_{211}}{k^4}  
+ \frac{1}{(n-1)} ( 2A(m_4^2) B(m^2,m^2,0) + m_3^2 P^{00}_{211}
\nonumber\\&&
+ \frac{1}{k^2} ( - (k^2 +m^2+m_3^2-m_4^2) P^{01}_{211} 
          - P^{01}_{111}  - 3P^{02}_{211}  + 
         m_3^2 P^{10}_{211}  - 2P^{11}_{211}  
\nonumber\\&&
- 3\frac{ P^{12}_{211}}{k^2} ) ) \Big)
\Big\}_y [m^2,m^2_3,m^2_4;k^2]\,,
\ea
where 
\be
\{ \ldots \}_y  \to \int^1_0\, dy \{\ldots \}\,,
\ee
and we have display the $P^{ij}_{abc}$ arguments outside the brackets.

The functions $P^{ij}_{111}$ can be obtained in terms of the $P^{ij}_{211}$ via recursion
relations. The reason of choosing $P^{ij}_{211}$ as a \emph{basis} instead of the 
simpler $P^{ij}_{111}$ is the infra-red behaviour in the last set. As is evident
the $P^{ij}_{111}$ have a direct relation with the $H_i$ functions 
defined in \cite{ABT1}, and we have checked that they agree with each other.
The relations between the $P^{ij}_{klm}$ functions with the $9$ $h_i$ is given in
Minkowsky space-time by (see \cite{GY})
\ba
 && (16 \pi^2)^2  P^{00}_{111}(1,2,3;k^2) =
        \frac{\lambda_2}{2} ( m_1^2 + m_2^2 + m_3^2 )
\nonumber\\&& \hspace{2.5cm}
       + \lambda_1 \Big(  - \frac{k^2}{4} 
       +  m_1^2(  \frac{1}{2} - \ln_1 ) + m_2^2 ( \frac{1}{2} - \ln_2) 
         +m_3^2 ( \frac{1}{2}  - \ln_3) \Big)
\nonumber\\&& \hspace{2.5cm}
       + k^2 \Big(  - \frac{9}{8} + \frac{1}{2}\ln_1 - h_{1;123} + 
         h_{2;123} \Big)
       +  m_1^2 ( \frac{1}{2} +\frac{\pi^2}{12} + \ln_1^2 - \ln_1 
         + h_{1;123} )  
\nonumber\\&& \hspace{2.5cm}
        + m_2^2 ( \frac{1}{2} +\frac{\pi^2}{12}+ \ln_2^2 - \ln_2 + h_{1;213} ) 
        + m_3^2 (\frac{1}{2}+\frac{\pi^2}{12}+ \ln_3^2 - \ln_3 
          + h_{1;312} )\,,
\nonumber\\
 && (16 \pi^2)^2  P^{01}_{111}(1,2,3;k^2) =
        -\frac{\lambda_2}{4}k^2 ( m_1^2 + m_3^2 )
\nonumber\\&& \hspace{2.5cm}
       - \frac{\lambda_1}{2} k^2 \Big(  - \frac{k^2}{12}
        + m_1^2 (\frac{1}{4} - \ln_1) + \frac{m_2^2}{2}
          +m_3^2(\frac{1}{4} - \ln_3 ) \Big)
\nonumber\\&& \hspace{2.5cm}
       - k^4 \Big(  - \frac{3}{8} + \frac{1}{6}\ln_2 - \frac{2}{3}h_{1;213}
         + \frac{7}{6}h_{2;213} - \frac{1}{2}h_{4;213} \Big)
\nonumber\\&& \hspace{2.5cm}
       - k^2 \Big( m_1^2 ( -\frac{3}{16} +\frac{\pi^2}{24} + \frac{1}{2}\ln_1^2 - \frac{1}{4}
         \ln_1 + \frac{1}{6}h_{1;123}+ \frac{2}{3}h_{3;123} )
\nonumber\\&& \hspace{2.5cm}
- m_2^2 (\frac{7}{8} - \frac{1}{2}\ln_2+ \frac{2}{3}(h_{1;213}-h_{2;213}) ) 
\nonumber\\&& \hspace{2.5cm}
         - m_3^2 ( -\frac{3}{16} + \frac{\pi^2}{24} + 
         \frac{1}{2}\ln_3^2 - \frac{1}{4}\ln_3  
         + \frac{1}{6}h_{1;321}
         + \frac{2}{3} h_{3;321}
          ) \Big)\,,
\nonumber\\
 && P^{10}_{111}(1,2,3;k^2) = P^{01}_{111}(2,1,3;k^2)\,,
\nonumber\\
 && (16 \pi^2)^2  P^{00}_{211}(1,2,3;k^2) =
\frac{1}{2} \lambda_2
- \frac{1}{2} \lambda_1 (1 + \ln_1) 
- \frac{1}{2} + \frac{\pi^2}{12}  + \ln_1^2 + \ln_1 
+ h_{1;123}\,, 
\nonumber\\
 && (16 \pi^2)^2   P^{10}_{211}(1,2,3;k^2) =
- \frac{\lambda_1}{4} k^2
       + k^2 \Big(  - \frac{5}{8} + \frac{1}{2} \ln_1 - h_{1;123} + 
         h_{2;123} \Big)\,,
\nonumber\\ 
 && (16 \pi^2)^2  P^{01}_{211}(1,2,3;k^2) =
 - \frac{\lambda_2}{4}  k^2
+ k^2 \lambda_1 ( \frac{3}{8} + \frac{1}{2} \ln_1 )
\nonumber\\&&\hspace{2.5cm}
+ \frac{k^2}{2} \Big( \frac{9}{8} - \frac{\pi^2}{12} - \ln_1^2 - \frac{3}{2}\ln_1  
-2 h_{3;123} \Big)\,,
\nonumber\\  
 && (16 \pi^2)^2   P^{20}_{211}(1,2,3;k^2) = 
 \frac{\lambda_2}{4} k^2 ( m_1^2 + \frac{1}{2} m_2^2 + \frac{1}{2} m_3^2 )
\nonumber\\&&\hspace{2.5cm}
 + \lambda_1 k^2 \Big(  \frac{m_1^2}{2} (\frac{1}{4} - \ln_1) + \frac{1}{4} m_2^2 (
       \frac{3}{4} - \ln_2) 
+ \frac{1}{4} m_3^2 (\frac{3}{4} - \ln_3)  \Big)
\nonumber\\&&\hspace{2.5cm}
+k^4 \Big(  - \frac{1}{6} + \frac{1}{2}h_{1;123} - \frac{5}{4}h_{2;123}
         + \frac{3}{4}h_{4;123} \Big)
\nonumber\\&&\hspace{2.5cm}
       + k^2 \Big( \frac{m_1^2}{2} ( \frac{1}{8} + \frac{\pi^2}{12}  + \ln_1^2 - \frac{1}{2}
         \ln_1 )  + \frac{m_2^2}{4} ( \frac{7}{8}+\frac{\pi^2}{12} + \ln_2^2 - \frac{3}{2}
         \ln_2 ) 
\nonumber\\&&\hspace{2.5cm}
+ \frac{m_3^2}{4} ( \frac{7}{8}+\frac{\pi^2}{12} + \ln_3^2 - \frac{3}{2} 
         \ln_3 ) + \frac{1}{2} m_1^2 h_{1;123}  + \frac{1}{4} m_2^2 h_{1;213}
         + \frac{1}{4} m_3^2 h_{1;312} \Big)\,,
\nonumber\\  
 && (16 \pi^2)^2   P^{11}_{211}(1,2,3;k^2) =
       -\frac{\lambda_2}{8}  k^2(  m_1^2 + m_3^2 )
\nonumber\\&&\hspace{2.5cm}
+\frac{\lambda_1}{4} k^2  \Big( \frac{k^2}{2}
       + m_1^2 (\ln_1 - \frac{1}{4}) - \frac{m_2^2}{2}
          + m_3^2 (\ln_3 - \frac{m_3^2}{4} ) \Big)
\nonumber\\&&\hspace{2.5cm}
       + k^4 \Big( \frac{19}{48} - \frac{1}{4}\ln_1 + \frac{1}{4} h_{1;123}
          - \frac{3}{8} h_{2;123} + h_{3;123} - \frac{3}{4} h_{5;123}
          \Big)
\nonumber\\&&\hspace{2.5cm}
       + k^2 \Big(  - \frac{m_1^2}{4} ( \frac{1}{8} + \frac{\pi^2}{12}+ \ln_1^2 - \frac{1}{2}
         \ln_1 + h_{1;123} ) 
\nonumber\\&&\hspace{2.5cm}
+\frac{m_2^2}{4} ( \frac{1}{4} - \frac{1}{4}\ln_1^2 
+ \frac{1}{2}\ln_1\ln_2
          - \frac{1}{4}\ln_2^2 + \ln_2 
- \frac{1}{2} (h_{1;123} + h_{1;213}) )
\nonumber\\&&\hspace{2.5cm}
+\frac{m_3^2}{8} ( -\frac{9}{4}  - \frac{\pi^2}{6}
         + \frac{1}{2}\ln_1^2 - \ln_1\ln_3 - \frac{3}{2}
         \ln_3^2 + \ln_3  
          + h_{1;123}
          - h_{1;312}) \Big)\,,
\nonumber\\  
 && (16 \pi^2)^2   P^{02}_{211}(1,2,3;k^2) =
      \frac{\lambda_2}{4} k^2 \Big( \frac{k^2}{2}
       + \frac{m_2^2}{2}  + m_3^2 \Big)
       - \frac{\lambda_1}{4} k^2 \Big(  k^2 (\frac{11}{12}  + \ln_1 )
       +  m_2^2 ( \frac{1}{4} +\ln_1 ) 
\nonumber\\&&\hspace{2.5cm}
       +  m_3^2 ( \frac{1}{2} +\ln_1 + \ln_3 ) \Big)
       + \frac{k^4}{4} \Big(  - \frac{37}{24} + \frac{\pi^2}{12} + \ln_1^2 + \frac{11}{6}
         \ln_1 + 3 h_{6;123} \Big)
\nonumber\\&&\hspace{2.5cm}
       + \frac{k^2}{4} \Big( m_2^2( - \frac{5}{8} + \frac{\pi^2}{12} + \ln_1^2 
         + \frac{1}{2}\ln_1 + h_{1;123} ) 
\nonumber\\&&\hspace{2.5cm}
+       m_3^2 ( \frac{3}{4} + \frac{\pi^2}{6} + \frac{1}{2} \ln_1^2 + 
         \ln_1\ln_3 + \frac{1}{2}\ln_1 + \frac{1}{2}\ln_3^2 + \frac{1}{2}
         \ln_3 )   \Big)\,,
\nonumber\\  
 &&(16 \pi^2)^2  P^{30}_{211}(1,2,3;k^2) =
       - \frac{\lambda_2}{8} k^4 (  m_2^2 + m_3^2 )
\nonumber\\&&\hspace{2.5cm}
       +\frac{\lambda_1}{4} k^4 \Big( \frac{k^2}{12}
         - m_1^2 + m_2^2 ( \ln_2 - \frac{5}{12})
         + m_3^2 (\ln_3 - \frac{5}{12} ) \Big)
\nonumber\\&&\hspace{2.5cm}
       + \frac{k^4}{2} \Big( k^2 (\frac{29}{96} - \frac{1}{12}\ln_1 - \frac{1}{3}h_{1;123}
         + \frac{11}{6}h_{2;123} - \frac{5}{2}h_{4;123} + 
         h_{7;123} )
\nonumber\\&&\hspace{2.5cm}
       +  m_1^2 (- \frac{5}{3}+\ln_1 - \frac{5}{3}(h_{1;123}- h_{2;123} ) )
\nonumber\\&&\hspace{2.5cm}
        + m_2^2 ( \frac{17}{144} - \frac{\pi^2}{24} - 
         \frac{1}{2}\ln_2^2 + \frac{5}{12}\ln_2 - \frac{1}{6}(h_{1;213}+ \frac{2}{3}h_{3;213})) 
\nonumber\\&&\hspace{2.5cm}
      +m_3^2 ( \frac{17}{144}    - \frac{\pi^2}{24}
          - \frac{1}{2}\ln_3^2 + \frac{5}{12}\ln_3  
         - \frac{1}{6} h_{1;312}
          - \frac{2}{3} h_{3;312}
          )
           \Big)\,,
\nonumber\\  
 &&(16 \pi^2)^2  P^{21}_{211}(1,2,3;k^2) =
        -\frac{\lambda_2}{8}  k^4 m_1^2
       + \frac{\lambda_1}{4} k^4 \Big(  - \frac{k^2}{24}
       + m_1^2 ( \frac{1}{4} + \ln_1 ) - \frac{m_3^2}{3} \Big)
\nonumber\\&&\hspace{2.5cm}
       + k^6 \Big( \frac{1}{128} + \frac{5}{48} \ln_1 - \frac{1}{12}\ln_2 - \frac{1}{4}h_{1;123}
         + \frac{1}{3}h_{1;213} + \frac{3}{8}h_{2;123}
         - \frac{7}{12}h_{2;213} 
\nonumber\\&&\hspace{2.5cm}
- \frac{1}{2}h_{3;123} + 
         \frac{1}{4}h_{4;213} + \frac{3}{4}h_{5;123} - \frac{1}{2}h_{8;123}
          \Big)
       + k^4 \Big(  m_1^2 ( \frac{37}{96}- \frac{\pi^2}{48} - \frac{1}{4}\ln_1^2 
\nonumber\\&&\hspace{2.5cm}
- \frac{1}{8}
         \ln_1 + \frac{1}{6} (h_{1;123}- 5h_{3;123} ) )
+ m_2^2( -\frac{7}{12} +\frac{1}{16}\ln_1^2 - \frac{1}{8}\ln_1
         \ln_2 + \frac{1}{16}\ln_2^2 
\nonumber\\&&\hspace{2.5cm}
- \frac{5}{24}h_{1;213}+ \frac{1}{3}h_{2;213}
+\frac{1}{8}h_{1;123})
+m_3^2 ( \frac{11}{36}- \frac{1}{16}\ln_1^2 + \frac{1}{8}
         \ln_1\ln_3 - \frac{1}{16}\ln_3^2 
\nonumber\\&&\hspace{2.5cm}
+ \frac{1}{6}\ln_3 
         - \frac{1}{8}h_{1;123} 
         + \frac{1}{24}h_{1;312} 
         - \frac{1}{3}h_{3;321}
         )
          \Big)\,,
\nonumber\\  
 &&(16 \pi^2)^2  P^{12}_{211}(1,2,3;k^2) =
        \frac{\lambda_2}{8} k^4 ( m_1^2 + m_3^2 )
\nonumber\\&&\hspace{2.5cm}
       + \frac{\lambda_1}{4} k^4 \Big(  - \frac{k^2}{4} 
        + m_1^2 ( \frac{1}{12} - \ln_1 ) + m_3^2 ( \frac{1}{12} - \ln_3 )
          \Big)
\nonumber\\&&\hspace{2.5cm}
       + k^6 \Big(  - \frac{43}{192} + \frac{7}{72}\ln_1 + \frac{1}{36}\ln_2 - \frac{1}{12}
         h_{1;123} - \frac{1}{9}h_{1;213} + \frac{1}{8}h_{2;123}
         + \frac{7}{36}h_{2;213} 
\nonumber\\&&\hspace{2.5cm}
- \frac{1}{6}h_{3;123} 
         -\frac{1}{24}h_{4;213} + \frac{1}{4}h_{5;123} - \frac{3}{4}h_{6;123}
         + \frac{1}{2}h_{9;123} \Big)
       + k^4 \Big( m_1^2 ( -\frac{31}{288} +\frac{\pi^2}{48}
\nonumber\\&&\hspace{2.5cm}
+ \frac{1}{4}\ln_1^2 - \frac{1}{24}
         \ln_1 + \frac{1}{9}h_{1;123} + \frac{5}{18} h_{3;123} )
+ \frac{m_2^2}{6} ( -\frac{1}{8} + \frac{1}{8} \ln_1^2 - \frac{1}{4}\ln_1
         \ln_2 
\nonumber\\&&\hspace{2.5cm}
+ \ln_1 + \frac{1}{8}\ln_2^2 - \ln_2  - \frac{5}{4}h_{1;123} 
+ \frac{11}{12}h_{1;213}
+ h_{2;123} - \frac{2}{3}h_{2;213}
+ h_{3;123})
\nonumber\\&&\hspace{2.5cm}
+ m_3^2 ( \frac{77}{288} + \frac{\pi^2 }{48} - \frac{1}{16}\ln_1^2 + \frac{1}{8}
         \ln_1\ln_3 
+ \frac{3}{16}\ln_3^2 - \frac{1}{24}\ln_3 
\nonumber\\&&\hspace{2.5cm}
         - \frac{1}{24}h_{1;123} 
         + \frac{5}{72}h_{1;312} 
         + \frac{1}{9}h_{3;321} 
         - \frac{1}{6}h_{3;123}
	 ) \Big)\,.
%
%
\ea
Where for sake of brevity the following notation should be understood
\ba
P^{ij}_{klm}(a,b,c,k^2)& =& P^{ij}_{klm}(m_a^2,m_b^2,m_c^2,k^2)\,,\nonumber\\ 
h_{i;abc}& =& h_i(m_a^2,m_b^2,m_c^2,k^2)\,,\nonumber\\
\ln_a &=& \ln(\frac{m_a^2}{\mu^2})\,,\nonumber\\
\lambda_2 &=& \lambda_0^2+(\ln(4\pi)+1-\gamma)^2\,,\nonumber\\
\lambda_1 &=& \lambda_0+\ln(4\pi)+1-\gamma\,,
\ea
and $\lambda_0$ is given in Eq. (\ref{pole}).
We have also employed some direct relations between the integral representation of the $h_i$
functions
\be
h_{1;abc} = h_{1;acb}\,,~~ \mbox{{and}} \quad h_{3;123}+h_{3;132} = h_{2;123}\,.
\ee


\end{document}